\documentclass[10pt,notitlepage,aps,pre,superscriptaddress]{revtex4-2}

\usepackage{amsfonts}

\usepackage{graphicx}
\usepackage{amsmath}
\usepackage{color}
\usepackage{titlesec}
\usepackage{footnote}
\usepackage{color,soul}
\definecolor{lightblue}{rgb}{.90,.95,1}
\sethlcolor{lightblue}

\usepackage{xcolor}
\usepackage{tcolorbox}

	\usepackage[colorlinks,hyperindex]{hyperref}
 \usepackage{hyperref}
	\hypersetup
	{
		colorlinks,%
		citecolor=blue,%
		linkcolor=blue,%
		urlcolor=blue,%
	}

\begin{document}

\title{Fluttering induced flow in a closed chamber}

\author{Kirill Goncharuk} 
%\email{goncharuk.kirill@gmail.com} 
\affiliation{Department of Mechanical Engineering, Ben-Gurion University of the Negev, Beer-Sheva 84105, Israel}

\author{Yuri Feldman} 
%\email{feldy77@gmail.com}
\affiliation{Department of Mechanical Engineering, Ben-Gurion University of the Negev, Beer-Sheva 84105, Israel}

\author{Oz Oshri} 
\email{oshrioz@bgu.ac.il} 
\affiliation{Department of Mechanical Engineering, Ben-Gurion University of the Negev, Beer-Sheva 84105, Israel}

\date{\today}
\begin{abstract}
We study the emergence of fluid flow in a closed chamber that is driven by dynamical deformations of an elastic sheet. The sheet is compressed between the sidewalls of the chamber and partitions it into two separate parts, each of which is initially filled with an inviscid fluid. When fluid exchange is allowed  between the two compartments of the chamber, the sheet becomes unstable, and its motion displaces the fluid from rest. We derive an analytical model that accounts for the coupled, two-way, fluid-sheet interaction. We show that the system depends on four dimensionless parameters: the normalized excess length of the sheet compared to the lateral dimension of the chamber, $\Delta$; the normalized vertical dimension of the chamber; the normalized initial volume difference between the two parts of the chamber, $v_{\text{du}}(0)$; and the structure-to-fluid mass ratio, $\lambda$. We investigate the dynamics at the early times of the system's evolution and then at moderate times. We obtain the growth rates and the frequency of vibrations around the second and the first buckling modes, respectively. Analytical solutions are derived for these linear stability characteristics within the limit of the small-amplitude approximation. At moderate times, we investigate how the sheet escapes from the second mode. Given the chamber's dimensions, we show that the initial energy of the sheet is mostly converted into hydrodynamic energy of the fluid if $\lambda\ll 1$, and into kinetic energy of the sheet if $\lambda\gg 1$. In both cases most of the initial energy is released at time $ t_{\text{p}}\simeq \ln[c \Delta^{1/2}/v_{\text{du}}(0)]/\sigma$, where $\sigma$ is the growth rate and $c$ is a constant.  
\end{abstract}

\maketitle
\vspace{1pt}

%\begin{keywords}
%\end{keywords}

%{\bf MSC Codes }  {\it(Optional)} Please enter your MSC Codes here

\section{Introduction}
\label{sec:intro}

Many natural processes and technological applications rest on fluid-structure interactions to maintain their regular functionality. Of particular interest are the mutual interactions between slender elastic objects and a fluid medium that trigger elasto-hydrodynamic instabilities. Such instabilities are vital for the control, for example, of the passage of air through the lungs \cite{Grotberg2004, Ishizaka1972}, the directionality of blood flow \cite{Pedley1996}, and the blood pressure of tall animals \cite{Pedley1996}. Moreover,  bending deformations of slender objects in viscous or inertial fluids have been manipulated for applications in soft robotics  \cite{Kim2013,Rothemund2018,Matia2015},  the fabrication of microfluidic soft actuators \cite{Thorsen2002,boyko2019,Holmes2013,Gomez2017,Liu2021,Hosoi2004,Neukirch2014,christov2018}, the manufacture of semiconductors \cite{king1989}, and the design of soft and active matter through catalytic reactions \cite{Laskar2022,Manna2022} and dynamical wrinkles \cite{Pocivavsek2019,Kodio2017,Chopin2017,Diamant2021,guan2022compression,doi:10.1073/pnas.1905755116,PhysRevFluids.5.014003,GUAN2023112242}.

Despite recent achievements, novel designs of small-scale devices still call for a deeper understanding of elasto-hydrodynamic couplings. One such design was recently introduced by Oshri~\cite{Oshri2021}. In that setup, a thin sheet is compressed between the two sides of a closed chamber and divides it into two separate parts that are connected by a valve (figure~\ref{schematics}). At time $t< 0$, the valve is closed, and each part of the chamber is filled with an incompressible fluid. In the absence of fluids, the sheet would have accommodated its minimum energetic state, i.e., the lowest mode of buckling, but in the presence of fluids, the sheet is forced to accommodate a higher energetic state. The additional energy can be exploited to displace the fluid from rest, if, for example, the valve is opened to allow the transfer of fluids between the two compartments of the chamber. 

In the above mentioned study, Oshri~\cite{Oshri2021} analyzed the quasi-static evolution of the system, wherein the volume of fluid exchanged between the two parts of the chamber is the control parameter. In contrast, the present work focuses on the dynamical evolution of the system, wherein the fluid is driven by the spontaneous relaxation of the sheet from higher to lower energetic states. We believe that the dynamic analysis of this setup will open new avenues for designing advanced technological devices, such as micro-mechanical switches \cite{Zhang2014,Preston2019,Krylov_2008} and microfluidic mixing devices \cite{Lee2011,Stroock2002,Liu2004}.  Indeed, the additional coupling between the sheet and the surrounding fluid confers increased flexibility in the design of such switches. Different fluids with different viscosities can be used to manipulate the time that required for the sheet to release its stored energy, thereby increasing, for example, the timescales over which the switches operate. In addition, when the two parts of the chamber are filled with different fluids, the elastic energy released from the sheet can be exploited for mixing: The pressure field induced in the chamber can be utilized to inject the fluid from one side of the chamber into the fluid on the other side, thereby inducing mixing of the two fluids. 
Typically, such devices function in conditions of low Reynolds numbers, where the effects of viscosity are significant. However, our system can also be applied in the design of pneumatic time-delay switches and soft pneumatic actuators \cite{Rothemund2018,Preston2019,Dylan2021}, which typically operate in the opposing limit of high Reynolds numbers.

%{\color{blue}{The latter devices usually operate in low Reynolds numbers, where viscose effects are important. However, our setup can also find applications in the design of pneumatic time-delay switches and soft pneumatic actuators \cite{Rothemund2018,Preston2019,doi:10.1126/scirobotics.aay2627},  which usually operate in the opposite limit of high Reynolds numbers.  }}

While successful implementation of these applications is in itself a challenging task (which we plan to pursue in future research), in this work, we aim to answer more fundamental questions related to the underlying physical behaviour of the system. For example, how much of the initial elastic energy is subsequently transferred from the sheet to the fluid? How is the velocity of the fluid that is induced in the chamber related to the elastic properties of the sheet? What is the maximum pressure difference that the sheet induces in the chamber?

As a first step to answering these questions, we derive an analytical model that encompasses the elasticity of thin sheets and the hydrodynamics of inviscid fluids. Our model reveals that the system depends on four dimensionless parameters:  the normalized excess length of the sheet compared to the lateral dimension of the chamber, $\Delta$, where the total length of the sheet is used to normalize all lengths; the normalized vertical dimension of the chamber, $L_y$; the normalized initial volume difference in the chamber, $v_{\text{du}}(0)$; and the structure-to-fluid mass ratio, $\lambda$.  We show that for fixed dimensions of the chamber, $L_y$ and $\Delta$, the system exhibits two asymptotic solutions as a function of $\lambda$. The sheet's inertia dominates the dynamics when $\lambda \gg 1$, and is therefore referred to below  as the ``solid-dominated'' region,  while the dynamics is governed by the fluid's inertia when $\lambda\ll 1$, and is therefore referred to as the ``fluid-dominated'' region.

We investigate the system's behaviour both in the early stages of its evolution and at moderate times during which nonlinear effects control the dynamics. For the early stages, we employ linear stability analysis around the (unstable) second buckling mode and the (stable) first buckling mode. We obtain the highest growth rate, $\sigma$, and the lowest frequency of vibration, $\omega$, around these initial states. The two solutions exhibit similar behaviour as a function of $\lambda$, namely, they converge to a constant in the solid-dominated region, while they exhibit the scaling $\lambda^{1/2}$  in the fluid-dominated region. Furthermore, we show that  in the solid-dominated region only one mode of the sheet is essentially excited at the instability, while an infinite number of modes are excited in the fluid-dominated region. Analytical approximations are derived for each of these cases under the assumption that the amplitude of the sheet remains small, i.e.,  $\Delta\ll 1$ \cite{Landau}. 

At moderate times, the weakly nonlinear analysis is performed around the second buckling mode. Given a small initial volume difference between the upper and lower parts of the chamber, we analyze the dynamic evolution of the system up to the peak time $t_{\text{p}}$, at which the sheet releases most of its initial potential energy. We show that, after some initial delay, the sheet rapidly escapes from the unstable state. We derive the approximation $\sigma t_{\text{p}}\simeq \ln \left[c\Delta^{1/2}/v_{\text{du}}(0)\right]$, where $\sigma$ is the growth rate of the linear instability and $c$ is a constant, and show that it agrees well with the numerical results.  At $t_{\text{p}}$, most of the initial potential energy is converted into a kinetic energy of the sheet if $\lambda\gg 1$, and into a hydrodynamic energy if $\lambda\ll 1$. We show that at $t=t_{\text{p}}$ relatively large spike of pressure drop is applied on the sheet.  

The paper is organized as follows. In $\S$~\ref{formulation}, we first formulate the problem for finite excess lengths. Then, we reduce this formulation to the small-amplitude approximation and introduce the modal expansion of the solution. In $\S$~\ref{early-time}, we investigate the early stages of the evolution. After recalling the static solution, we employ a linear stability analysis around the second and the first modes of buckling. In $\S$~\ref{moderate-time}, we investigate the system's evolution at moderate times. In particular, we examine the energetic interplay between the sheet and the fluid, derive the scaling for the peak time, $t_{\text{p}}$, and explore the relation between the volume difference and the pressure drop on the sheet. Finally, in $\S$~\ref{discussion}, we discuss a possible experimental realization of the system, and in $\S$~\ref{conclusions} we draw conclusions, and propose a direction for future study.

\section{Formulation of the problem}
\label{formulation}
\begin{figure}[h!]
\begin{centering}
\includegraphics[height=6cm]{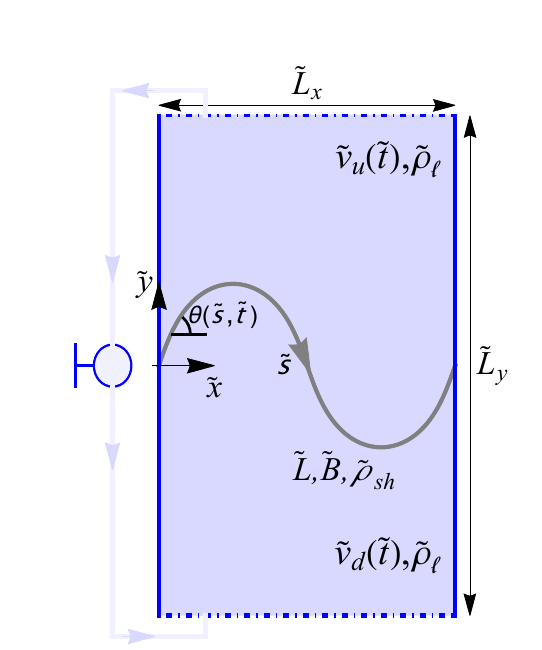} 
%\captionsetup{format=plain,justification=justified}
\caption{Schematic overview of the system. A thin sheet of total length $\tilde{L}$, bending modulus $\tilde{B}$, density $\tilde{\rho}_{\text{sh}}$, and thickness $\tilde{h}$ divides a closed rectangular chamber of dimensions $\tilde{L}_x\times \tilde{L}_y$ into two parts. %that are connected by a valve. 
The excess length of the sheet compared to the lateral dimension of the chamber is given by $\tilde{\Delta}=\tilde{L}-\tilde{L}_x$ (not shown in the figure). The volumes of the chamber above and below the sheet, $\tilde{v}_{i}(t)$ ($i=$u,d),  are filled with an inviscid and irrotational fluid of density $\tilde{\rho}_{\ell}$.  At $\tilde{t}\geq 0$, fluid is allowed to exchange freely between the two compartments of the chamber. In our formulation, the fluid exchange occurs through the upper and lower walls of the chamber (represented by dashed-dotted blue lines). To model this exchange, we apply periodic boundary conditions along these walls. One possible experimental setup that corresponds to the above model involves a valve-controlled channel that connects the two compartments of the chamber. %The design of possible experimental setup corresponding to the above model can, for example, utilize a valve controlled channel connecting both compartments of the chamber. %An experimental approximation of this fluid exchange can be achieved by utilizing a valve and a channel, as depicted schematically in the figure. 
%the valve is opened, and the fluid is allowed to exchange freely between the two parts of the chamber.
}
\label{schematics}
\end{centering}
\end{figure}

We consider an inextensible thin sheet of total length $\tilde{L}$, bending modulus $\tilde{B}$, thickness $\tilde{h}$, and density $\tilde{\rho}_{\text{sh}}$. %{\color{blue}{The inextensibility constraint in the elastic model implies that the sheet can only experience bending deformations, but not stretching deformations.}} 
The sheet divides a rectangular closed chamber into two parts, which are connected by a valve (figure~\ref{schematics}).    The lateral, the vertical, and  the width dimensions of the chamber are denoted by $\tilde{L}_x$, $\tilde{L}_y$, and $\tilde{W}$, respectively. A Cartesian coordinate system is located on the left edge of the sheet. A cross-section of the chamber on the $\tilde{x}\tilde{y}$ plane is placed at $0\leq \tilde{x}\leq \tilde{L}_x$ and $-\tilde{L}_y/2 \leq \tilde{y}\leq \tilde{L}_y/2$.  When $\tilde{t}<0$,  the valve connecting the two parts of the chamber is closed, and the volumes above and below the sheet, $\tilde{v}_{\text{u}}(\tilde{t})$ and $\tilde{v}_{\text{d}}(\tilde{t})$, are filled with an incompressible, inviscid fluid of density $\tilde{\rho}_{\ell}$. Hereafter, we denote quantities related to the upper and lower parts of the chamber by the subscripts `u' and `d', respectively. At $\tilde{t}\geq 0$,  the valve is opened, and free exchange of fluid is allowed in the chamber.

In the analysis that follows, we normalize all lengths by the total length of the sheet, $\tilde{L}$, and we normalize time by the inertial time-scale of the sheet  $\tilde{t}_{\star}=\tilde{L}^2(\tilde{\rho}_{\text{sh}} \tilde{h}/\tilde{B})^{1/2}$, i.e., 
\begin{equation}\label{normalization}
t=\tilde{t}/\tilde{t}_{\star}, \ \ \ \ \ \ x=\tilde{x}/\tilde{L}, \ \ \ L_x=\tilde{L}_x/\tilde{L}, \ \ \ \ v_{\text{d}}(t)=\tilde{v}_{\text{d}}(\tilde{t})/\tilde{L}^3, \ \ \ \text{etc.} 
\end{equation}

We choose this normalization because we anticipate that the wavelengths on the sheet will scale with the sheet's total length. In addition, since the dynamics in the system are driven by the sheet's motion, we chose the sheet's inertial timescale for the normalization. Note that our normalization with respect to lengths and time implies the normalization of the hydrodynamic fields and of the elastic fields, as will be emphasized further during the formulation. Hereafter, we denote all dimensional quantities with tilde over the symbol, and the corresponding nondimensional quantities without a tilde. 

Our model is based on the following six assumptions. Firstly, we assume that the system remains uniform along the width dimension of the chamber. Therefore, we set $W=1$ and consider a two-dimensional system. Secondly, we assume that the volume occupied by the elastic sheet is negligible compared to the total volume of the chamber, i.e., $\tilde{h}\tilde{L}/(\tilde{L}_x \tilde{L}_y) \ll 1$, and as a result $v_{\text{u}}(t)+v_{\text{d}}(t)=L_x L_y$. Thirdly, we assume that the fluid exchange between the two parts of the chamber occurs through the upper and lower walls, i.e., the walls located at $y=\pm L_y/2$. Fourthly, we assume that the vertical dimension of the chamber, $L_y$, is larger than the typical length scale, $\ell$, over which the disturbances in the flow caused by the sheet's motion decay to zero.
%Fourthly, we assume that the vertical dimension of the chamber, $L_y$, is larger than the typical length scale, $\ell$, over which the disturbances in the flow, due to the motion of the sheet, decay to zero.}} %Our analysis shows that this decay length scales with the total length of the sheet, $\tilde{L}$, and depends on the dynamic mode number that dominates the sheet's motion.  
In addition, we assume that there is no contact between the sheet and the sidewalls of the chamber, or of the sheet with itself, at any time during the system's evolution. Lastly, we assume that at $t=0$ the system is at rest and that the sheet accommodates a configuration that is dictated by the volume difference $v_{\text{du}}(0)=v_{\text{d}}(0)-v_{\text{u}}(0)$.

For an inviscid and irrotational fluid, the state of the flow is determined by four fields. Two of these are the fluid's potential functions $\phi_i(x,y,t)$, where i$=$u,d, from which we can determine the velocity profile of the fluid as ${\bf v}_i=\nabla \phi_i$, where $\nabla$ is the two-dimensional  gradient operator. The other two fields that characterize the flow are the pressures $p_i(x,y,t)$ in each side of the chamber. Using our normalization convention, we find that the potential functions are normalized by $\phi_i=\tilde{\phi}_i(\tilde{\rho}_{\text{sh}} \tilde{h}/\tilde{B})^{1/2}$, and the pressures, by $p_i=\tilde{p}_i\tilde{L}^3/\tilde{B}$. The evolution of these hydrodynamic fields, in space and over time, is determined by the continuity equation and Bernoulli's equation:
\begin{subequations}\label{continuity-benoulli}
\begin{eqnarray}
\nabla^2 \phi_i&=&0, \label{continuity-benoulli-1}\\
\lambda p_{\text{i}}+\frac{\partial \phi_i}{\partial t}+\frac{1}{2}|\nabla \phi_i|^2 &=&c_i(t), \label{continuity-benoulli-2}
\end{eqnarray}
\end{subequations}
where $c_i(t)$ are arbitrary functions that depend on time. Throughout the system's development, these functions are employed to maintain a constant pressure at a point within each part of the chamber \cite{Lamb}. %These functions are used to fix the pressure at a point in each side of the chamber 
%At each moment in time 
%Throughout the system's evolution, these functions are used to maintain a constant pressure at a point in each part of the chamber \cite{Lamb}. 
%In the following analysis we determine $c_i(t)$ such that:
%\begin{equation}\label{ref-pressure}
%p_{\text{d}}\left(\frac{1-\Delta}{2},-\frac{L_y}{2},t\right)=p_{\text{u}}\left(\frac{1-\Delta}{2},\frac{L_y}{2},t\right)=0.
%\end{equation}
In addition, in Eqs.~(\ref{continuity-benoulli-2}) and (\ref{ref-pressure}), we define the dimensionless parameters:
\begin{equation}
\lambda=\frac{\tilde{\rho}_{\text{sh}} \tilde{h}}{\tilde{\rho}_{\ell}\tilde{L}}, \ \ \ \ \   \ \ \ \ \ \Delta=1-L_x.
\end{equation}
%\begin{subequations}
%\begin{eqnarray}
%\lambda&=&\frac{\tilde{\rho}_{\text{sh}} \tilde{h}}{\tilde{\rho}_{\ell}\tilde{L}}, \\
%\Delta&=&1-L_x,
%\end{eqnarray}
%\end{subequations}
%that are obtained from our normalization.
The structure-to-fluid mass ratio, $\lambda$, accounts for the ratio between the densities of the sheet and the fluid and the slenderness of the sheet. This dimensionless parameter plays a role, for example, in the problem of a flag flapping under a uniform axial flow \cite{connell2007,Argentina2005,alben_2008,Alben2008}. %This parameter will play an essential role in the characterization of the system, as is further clarified throughout our analysis.
The parameter $\Delta$ accounts for the difference between the total length of the sheet and the lateral dimension of the chamber. In dimensional form, it may be expressed as $\tilde{\Delta}=\tilde{L}-\tilde{L}_x$. For a given system, the parameters $\lambda$ and $\Delta$ remain constant throughout the dynamic evolution.

%, say $p_{\text{d}}((L-L_x)/2,-L_y/2,t)=0$.  %Following Ref.~\cite{Lamb}, these functions can be absorbed into the potential functions, $\phi_i(x,y,t)$, without altering the velocity fields. For this reason, in the following analysis we set $c_i(t)=0$. In addition, in Eq.~(\ref{continuity-benoulli-2}) we define the dimensionless parameter,
%\begin{equation}
%\lambda=\frac{\rho_{\text{sh}} h}{\rho_{\ell}L},
%\end{equation}
%that accounts for the ratio between the densities of the sheet and the fluid and the slenderness of the sheet. This dimensionless number plays a role, for example, in the problem of a flag flapping under a uniform axial flow \cite{connell2007,Argentina2005,alben_2008,Alben2008}. %This parameter will play an essential role in the characterization of the system, as is further clarified throughout our analysis.

To solve the continuity equation, Eq.~(\ref{continuity-benoulli-1}), we must first specify the boundary conditions on the chamber's walls and the fluid-sheet interfaces. Since the fluid that exits the upper wall of the chamber enters through the lower wall, we set periodic boundary conditions through $y=\pm L_y/2$. %Given that the flow is fully developed at the upper and lower boundaries, we start by setting the pressures on these walls to zero. 
Thereafter,  we ensure that there is no penetration of fluid through the sidewalls of the chamber. These restrictions give the boundary conditions:
\begin{subequations}\label{bc-fluid-chamber}
\begin{eqnarray}
\phi_{\text{u}}\left(x,\frac{L_y}{2},t\right)&=&\phi_{\text{d}}\left(x,-\frac{L_y}{2},t\right), \ \ \ \text{and} \ \ \ \frac{\partial\phi_{\text{u}}}{\partial y}\left(x,\frac{L_y}{2},t\right)=\frac{\partial\phi_{\text{d}}}{\partial y}\left(x,-\frac{L_y}{2},t\right), \nonumber \\   \label{bc-fluid-chamber-2}  \ \ \ %0<x<L_x,    
\label{bc-fluid-chamber-3}\\
\frac{\partial \phi_i}{\partial x}(0,y,t)&=&\frac{\partial \phi_i}{\partial x}(1-\Delta,y,t)=0.  \label{bc-fluid-chamber-1}
\end{eqnarray}
\end{subequations}
In addition to the periodic boundary conditions at $y=\pm L_y/2$, it is necessary to ensure that $p_{\text{u}}(x,L_y/2,t)=p_{\text{d}}(x,-L_y/2,t)$ along these walls. By utilizing Bernoulli's equation, Eq.~(\ref{continuity-benoulli-1}), and the periodic boundary conditions, it becomes apparent that this requirement is satisfied when $c_{\text{d}}(t)=c_{\text{u}}(t)\equiv c(t)$. Consequently, we can determine the function $c(t)$ by fixing the pressure at a specific point in the lower part of the chamber. In the following analysis we choose
%In addition to the periodic boundary conditions, Eq.~(\ref{bc-fluid-chamber-2}), we must also require that $p_{\text{u}}(x,L_y/2,t)=p_{\text{d}}(x,-L_y/2,t)$ along these walls. Using Bernoulli's equation, Eq.~(\ref{continuity-benoulli-1}), and the periodic boundary conditions, it is evident that this requirement is satisfied if $c_{\text{d}}(t)=c_{\text{u}}(t)$. We can now determine, for example, the function $c_{\text{d}}(t)$ if we fix the pressure at a point in the lower part of the chamber. For these reasons, in the following analysis we choose,
\begin{equation}\label{ref-pressure}
p_{\text{d}}\left(\frac{1-\Delta}{2},-\frac{L_y}{2},t\right)=0.
%p_{\text{d}}\left(\frac{1-\Delta}{2},-\frac{L_y}{2},t\right)=p_{\text{u}}\left(\frac{1-\Delta}{2},\frac{L_y}{2},t\right)=0.
\end{equation}

Two sets of equations model the contact between the sheet and the fluid. The first set corresponds to the kinematic boundary conditions that ensure continuous contact between the sheet and the fluid. The second set corresponds to the force balance equations on the sheet that ensure proper transfer of the momentum between the solid and the fluid. To obtain these two sets of equations, we first define the elastic fields that describe the position of the sheet on the $xy$ plane.  It is important to note that, as we assumed the sheet to be inextensible, the elastic model accounts only for bending deformations and does not include stretching deformations. 
In contrast to the Eulerian description of the fluid, it is convenient to adopt a Lagrangian description for the sheet and to define the elastic fields as functions of the normalized arclength parameter on the sheet, $s\in[0,1]$. With this change of reference frame, we define the position vector to a point on the sheet  as ${\bf x}_{\text{sh}}(s,t)=(x_{\text{sh}}(s,t),y_{\text{sh}}(s,t))$ and the angle between the tangent to the sheet and the $x$-axis as $\theta(s,t)$; see figure~\ref{schematics}. These three elastic fields, i.e.,  $x_{\text{sh}}(s,t)$, $y_{\text{sh}}(s,t)$ and $\theta(s,t)$,  are not independent, since they are related by the geometric constraints:
\begin{subequations}\label{geo-rel}
\begin{eqnarray}
&&\frac{\partial x_{\text{sh}}}{\partial s}=\cos\theta, \\ %x_{\text{sh}}(s,t)=\int_{0}^{s}\cos\theta(s',t)ds', \\
&&\frac{\partial y_{\text{sh}}}{\partial s}=\sin\theta.  %y_{\text{sh}}(s,t)=\int_{0}^{s}\sin\theta(s',t)ds'.
\end{eqnarray}
\end{subequations}
By using these definitions, the kinematic boundary conditions on the sheet-fluid interfaces are given by:
\begin{equation}\label{kinematic-bc}
\text{$y=y_{\text{sh}}(x(s),t)$}, \ \ \ \ \ \ \ \ \ \ \ \  \frac{D y_{\text{sh}}}{D t}=\frac{\partial \phi_i}{\partial y},
\end{equation}
where $D/Dt=\partial/\partial t +{\bf v}_i \cdot \nabla$ is the two-dimensional convective derivative. The balance of moments and forces on the sheet is given by:
\begin{subequations}\label{force-balance}
\begin{eqnarray}
%\text{$y=y_{\text{sh}}(x,t)$}, \ \ \ \ \ \ \ \ \ \ \ \  \frac{D y_{\text{sh}}}{D t}&=&\partial_y \phi_i, \\ %\ \ \ \ \  y=y_{\text{sh}}(x,t),  \label{kinematic-bc} \\
\frac{\partial^2\theta}{\partial s^2}&=&-F_x \sin\theta+F_y \cos\theta, \label{force-balance-1}\\
\frac{\partial^2 {\bf x}_{\text{sh}}}{\partial t^2}&=&-\frac{\partial {\bf F}}{\partial s} +\left[p_{\text{d}}(x_{\text{sh}},y_{\text{sh}},t)-p_{\text{u}}(x_{\text{sh}},y_{\text{sh}},t)\right]{\bf \hat{n}}_{\text{d}}, \label{force-balance-2} 
\end{eqnarray}
\end{subequations}
where  ${\bf F}=\left(F_x(s,t),F_y(s,t)\right)$ is the vector of reaction forces per unit length at a cross-section of the sheet, and our normalization implies that ${\bf F}=\tilde{{\bf F}}\tilde{L}^2/\tilde{B}$. In addition,  ${\bf \hat{n}}_{\text{d}}=(-\sin\theta,\cos\theta)$ is a unit normal vector on the sheet that points outwards from the lower part of the chamber, and the hydrodynamic pressures in Eq.~(\ref{force-balance-2}) are calculated on their respective sides of the sheet-fluid interfaces. Note that in  Eq.~(\ref{force-balance-1}) we neglect the rotational inertia term. This is justified in the limit of a thin and inextensible sheet, as assumed in this analysis \cite{Goriely,PhysRevE.101.053002,NEUKIRCH2012704}. %{\color{blue}{Note also that our normalization implies that ${\bf F}=\tilde{{\bf F}}\tilde{L}^2/\tilde{B}$ and that the arclength and the position vector are both normalized by the total length of the sheet, i.e., $s=\tilde{s}/\tilde{L}$ and ${\bf x}_{\text{sh}}=\tilde{{\bf x}}_{\text{sh}}/\tilde{L}$.}} 
Equations~(\ref{force-balance}) are supplemented by the following boundary conditions on the sheet's edges:
\begin{subequations}\label{bc-sheet}
\begin{eqnarray}
&&x_{\text{sh}}(0,t)=0, \ \ \ \ \ x_{\text{sh}}(1,t)= 1-\Delta,  \label{bc-sheet-1} \\
&&y_{\text{sh}}(0,t)=0,  \ \ \ \ \ y_{\text{sh}}(1,t)=0, \label{bc-sheet-2} \\
&&\frac{\partial \theta}{\partial s}(0,t)=0,\ \ \ \ \ \frac{\partial \theta}{\partial s}(1,t)=0, \label{bc-sheet-3} 
\end{eqnarray}
\end{subequations}
where we assume hinged boundary conditions in Eq.~(\ref{bc-sheet-3}). 

%So far, we introduced the formulation in dimensional form. Let us now scale all lengths in our model by the total length of the sheet, and to normalize time by the inertial time scale of the sheet,
%\begin{equation}
%\tilde{t}=t/t_{\star}, \ \ \ \ \ 
%\end{equation}

%In the analysis that follows, we normalize all lengths by the total length of the sheet, $L$, and we normalize time by the inertial time scale of the sheet, $L^2(\rho_{\text{sh}} h/B)^{1/2}$.

This completes the formulation of the problem. In summary, given the excess length $\Delta$, the vertical dimension of the chamber $L_y$, the parameter $\lambda$, and the initial volume difference in the chamber $v_{\text{du}}(0)$,  the dynamic evolution of the system is determined from the solution of the coupled equations~(\ref{continuity-benoulli})-(\ref{bc-sheet}). In the analysis that follows, we will always assume that the sheet and the fluid are initially at rest, i.e., $\frac{\partial{\bf x}_{\text{sh}}}{\partial t} (s,0)=0$ and $\phi_i(x,y,0)=0$. %, and the configuration of the sheet is a given function of the arclength parameter, i.e., ${\bf x}_{\text{sh}}(s,0)$ and $\theta(s,0)$. 

While solutions to our set of nonlinear equations can, in practice, be sought only numerically, some analytical progress that sheds light on the underlying physics of the system can be achieved under the assumption that the excess length remains small, i.e., $\Delta\ll 1$. For this reason, in the next section, we reduce our model to this so-called small-amplitude approximation \cite{Landau} and exploit this formulation to study the time-dependent behaviour of the system.
%, and so as the fluid's velocity. As we shall see below the later assumption is akin to the limit $\lambda\ll 1/\Delta$.  

 %that the amplitude of the sheet remains small compared with the lateral dimension of the chamber, and that the velocity of the fluid remains small. For this reason, in the next section we reduce our model to this so-called small amplitude approximation, and exploit this formulation to study the time dependent behavior of the system.

However, before we proceed to the next section, we should add a comment regarding the system’s energy. Since we assumed an ideal fluid, i.e., one without viscous dissipation, and since we consider an elastic model, our equations have a conserved first integral that corresponds to the system's total energy. In accordance with  Appendix~\ref{energy-der-app}, it can be shown that the total energy of the system is given by the sum of the energies of the sheet and the fluid,  $E=E_{\text{sh}}(t)+E_{\text{f}}(t)$, where $E_{\text{sh}}(t)$ accounts for the sum of the kinetic and the potential energies of the sheet, which are designated $E_{\text{sh}}^{\text{p}}(t)$ and $E_{\text{sh}}^{\text{k}}(t)$, respectively, and $E_{\text{f}}(t)$ accounts for  the kinetic energy of the fluid.  Therefore, the total energy is given by:
\begin{equation}\label{energy-tot}
E=\frac{1}{2}\int_0^1 \left[\left|\frac{\partial{\bf x}_{\text{sh}}}{\partial t}\right|^2+\left(\frac{\partial \theta}{\partial s}\right)^2\right]ds+\sum_{i=\text{u,d}}\frac{1}{2\lambda}\iint_{v_i(t)}|\nabla\phi_i|^2dx dy,
\end{equation}
where $|\  . \ |$ corresponds to the norm of the enclosed vector, and our normalization implies that $E=\tilde{E}\tilde{L}/\tilde{B}$. Since our system starts from rest, the total energy of the system equals the initial potential energy of the sheet, $E=E_{\text{sh}}^{\text{p}}(0)$, and this energy is conserved throughout the system's evolution. 

%{\color{blue}{ Note that in our normalization energy is normalized by $\tilde{E}=E\tilde{L}/B$.}}

\subsection{The small-amplitude approximation}
\label{small-slope-formulation}
The assumption that the amplitude of the sheet remains small, or equivalently that $\Delta\ll 1$, implies that the geometric relations, Eq.~(\ref{geo-rel}), reduce to $ \partial y_{\text{sh}}/\partial s\simeq \theta$ and $\partial x_{\text{sh}}/\partial s\simeq 1-\frac{1}{2}(\partial y_{\text{sh}}/\partial s)^2$. The non-linearity in the derivative of $x_{\text{sh}}(s,t)$ is retained in the leading order of the theory so as to satisfy the constraint on the excess length, Eq.~(\ref{bc-sheet-1}). Indeed, in the small-amplitude approximation, this constraint is given by:
\begin{equation}\label{constraint-app}
\Delta=\frac{1}{2}\int_0^1\left(\frac{\partial y_{\text{sh}}}{\partial x}\right)^2dx.
\end{equation}
Here, we replace the arclength coordinate of the sheet with the Eulerian coordinate of the fluid, $s\simeq x$, according to our level of approximation.  Correspondingly, the balance of forces and moments on the sheet, Eq.~(\ref{force-balance}),  reduces to:
\begin{equation}\label{force-balance-app}
\frac{\partial^2 y_{\text{sh}}}{\partial t^2}+\frac{\partial^4 y_{\text{sh}}}{\partial x^4}+F_x(t)\frac{\partial^2 y_{\text{sh}}}{\partial x^2}+\left[p_{\text{u}}(x,0,t)-p_{\text{d}}(x,0,t)\right]=0,
\end{equation}
where the lateral compression, $F_x(t)$, is a function that depends solely of time. In addition, the pressure difference that the fluid exerts on the sheet, i.e., the last term in Eq.~(\ref{force-balance-app}), is calculated at the sheet-fluid interface, $y=0$.

Thus far, we have approximated only the elastic part of the model. To further simplify the hydrodynamic part, we need to estimate the order of its corresponding fields. Given that the initial energy of the sheet scales linearly with the excess length,  $E_{\text{sh}}(0)\propto \Delta$, and that the total energy of the system is conserved,  the energy of the fluid is, at most,  proportional to $E_{\text{f}}\sim\Delta$. Therefore, if we approximate the energy of the fluid by $E_{\text{f}}\sim |{\bf{ v}}|^2 \ell$, where $|{\bf{ v}}|$ is the typical velocity in the chamber and $\ell$ is the decay length of the  disturbances in the flow, we obtain $|{\bf{ v}}|\sim \sqrt{\Delta/\ell}$. %Therefore, using Eq.~(\ref{energy-tot}), we expect the maximum velocity of the fluid at some instant in time to be proportional to $ \sqrt{\Delta}$. 
Furthermore, if we assume that the order of approximation of a derivative over the potential function, with respect to either  a spatial dimension or time, does not change, then we can approximate Bernoulli's equation, Eq.~(\ref{continuity-benoulli-2}), and the kinematic boundary conditions by:
\begin{subequations}\label{benuolli-kin}
\begin{eqnarray}
\lambda p_i+\frac{\partial\phi_i}{\partial t}&=&c(t),\label{benuolli-kin-1} \\
 \frac{\partial y_{\text{sh}}}{\partial t}&=&\left(\frac{\partial \phi_i}{\partial y}\right)_{y=0}. \label{benuolli-kin-2}
\end{eqnarray} 
\end{subequations} 
These approximations will further be verified {\it a posteriori} in $\S$~\ref{moderate-time}, where we analyze the nonlinear dynamics of the system. In particular, we will compare the results of this approximation with the numerical solution of the nonlinear model, Eqs.~(\ref{continuity-benoulli})-(\ref{bc-sheet}). Note that since the continuity equations, Eq.~(\ref{continuity-benoulli-1}), are already linear in the potential functions, they remain unchanged in our approximated model.

This completes the reduction of our model to the small-amplitude limit. In summary, Eqs.~(\ref{continuity-benoulli-1}) and (\ref{constraint-app})-(\ref{benuolli-kin}), supplemented by the linearized form of the boundary conditions, Eqs.~(\ref{ref-pressure}), (\ref{bc-fluid-chamber}), (\ref{bc-sheet-2}) and (\ref{bc-sheet-3}),  form a closure and describe the coupled dynamics of the sheet and the fluid in the small-amplitude approximation. A comment is necessary regarding this simplified formulation. %A comment is in order regarding this reduced formulation. 
In accordance with the derivation in Appendix~\ref{lag-mini}, it can be shown that the reduced model emanates from the minimization of the action $\mathcal{S}=\int_0^T\mathcal{L}dt$, where  
\begin{eqnarray}\label{lagrangian}
\mathcal{L}&=&\int_0^1\left[\frac{1}{2}\left(\frac{\partial y_{\text{sh}}}{\partial t}\right)^2-\frac{1}{2}\left(\frac{\partial^2y_{\text{sh}}}{\partial x^2}\right)^2+F_x(t)\left(\frac{1}{2}\left(\frac{\partial y_{\text{sh}}}{\partial x}\right)^2-\Delta\right)+\frac{1}{\lambda}\left[\phi_{\text{d}}(x,0,t)-\phi_{\text{u}}(x,0,t)\right]\frac{\partial y_{\text{sh}}}{\partial t}\right]dx%\nonumber \\ &-&\sum_{i=\text{u,d}}\frac{1}{2\lambda}\iint_{V_i}|\nabla\phi_i|^2dxdy,
\nonumber \\ &-&\frac{1}{2\lambda}\int_{0}^{\frac{L_y}{2}}\int_0^1|\nabla\phi_{\text{u}}|^2dxdy-\frac{1}{2\lambda}\int_{-\frac{L_y}{2}}^0\int_0^1|\nabla\phi_{\text{d}}|^2dxdy,
%\frac{1}{2\lambda}\int_{-\frac{L_y}{2}}^0\int_0^1|\nabla \phi_{\text{d}}|^2dxdy-\frac{1}{2\lambda}\int_0^{\frac{L_y}{2}}\int_0^1|\nabla\phi_{\text{u}}|^2dxdy.
\end{eqnarray}
with respect to the elastic fields $y_{\text{sh}}(x,t)$ and $F_x(t)$ and the hydrodynamic fields $\phi_{i}(x,y,t)$. In the next section, we employ a modal expansion of these fields and combine it with the Lagrangian formulation to derive a simplified set of equations that are dependent solely on time. 

%{\color{blue}{
%In the next section, we use a modal expansion to express the solution of the hydrodynamic potentials and the sheet's height function.  Then, we use the action, Eq.~(\ref{lagrangian}), to derive a close set of equations for the time-dependent coefficients of these expansions.}}

%Secondly, since our system starts from rest and since the boundary conditions on the upper and lower walls of the chamber require that $p_i(x,\pm L_y/2,t)=0$, we have from Eq.~(\ref{benuolli-kin-1}) that this condition is equivalent to $\phi_{i}(x,\pm L_y/2,t)=0$. 

\subsubsection{Modal expansion}

The continuity equations, Eq.~(\ref{continuity-benoulli-1}), and their corresponding boundary conditions on the  fluid-chamber interfaces, Eq.~(\ref{bc-fluid-chamber}), are satisfied when the potential functions are given by:
\begin{equation}\label{phi-exp}
\phi_i(x,y,t)=a_0(t)\left(y\pm \frac{L_y}{2}\right)+\sum_{m=1}^{\infty}\cos(\pi m x)\left[a_m(t)e^{\pi m\left(y\pm \frac{L_y}{2}\right)}+c_m(t)e^{-\pi m \left(y\pm\frac{L_y}{2}\right)}\right], 
\end{equation}
where $a_m(t)$ ($m=0,1,2,...$) and $c_m(t)$ ($m=1,2,3,...$) are unknown time-dependent coefficients, and the $\pm$ signs correspond to the solutions of the potential functions in the lower and the upper parts of the chamber,  respectively. Similarly, we expand the solution of the sheet's height function in the following normal modes:
\begin{equation}\label{ysh-exp}
y_{\text{sh}}(x,t)=\sum_{n=1}^{\infty}A_n(t)\sin(\pi n x),
\end{equation}
where the functions $\sin(\pi n x)$ automatically satisfy  the boundary conditions on the sheet edges, Eqs.~(\ref{bc-sheet-2}) and (\ref{bc-sheet-3}), and $A_n(t)$ are as-yet unknown coefficients. 

With these expansions, the solution to our problem reduces to finding the unknown coefficients, $a_m(t)$, $c_m(t)$ and $A_n(t)$, and the compression force $F_x(t)$, from the solution of the force balance equation, Eq.~(\ref{force-balance-app}), Bernoulli's equation, Eq.~(\ref{benuolli-kin-1}),  the kinematic boundary conditions, Eq.~(\ref{benuolli-kin-2}), and the geometric constraint, Eq.~(\ref{constraint-app}). Equation~(\ref{ysh-exp}) involves infinite summation over the modes of the height function, but, in practice, we will truncate this series at  $n=N$. A closed system of equations is then obtained when the coefficients of $a_m(t)$ and $c_m(t)$ are truncated at $N-1$.  

However, instead of directly using these equations, we take a different - yet equivalent - approach, by utilizing the Lagrangian formulation, Eq.~(\ref{lagrangian}). To this end, we follow the analysis in Appendix~\ref{lag-derivation} and substitute the potential functions, Eq.~(\ref{phi-exp}), and the height function, Eq.~(\ref{ysh-exp}), into the Lagrangian, Eq.~(\ref{lagrangian}). We then integrate over the spatial coordinates. Thereafter, we minimize the Lagrangian with respect to $a_m(t)$ and $c_m(t)$ and express these coefficients in terms of  $A_n(t)$. Substituting $a_m(t)$ and $c_m(t)$ back into the Lagrangian gives:
\begin{equation}\label{lag-app2}
\mathcal{L}\left[A_1,...,A_N,F_x\right]= T_{nk}\frac{dA_k}{dt}\frac{dA_n}{dt}-V_{nk}A_k A_n+F_x(t)C_{nk}A_k A_n-\Delta F_x(t),
\end{equation}
where  Einstein's summation rule is implied for repeated indices, and we define the following symmetric matrices:
\begin{subequations}\label{TVC-matrices}
\begin{eqnarray}
T_{nk}&=&\frac{1}{4}\delta_{nk}+\frac{L_y}{2\lambda }W(n,0)W(k,0)+\sum_{m=1}^{N-1}\frac{2}{\pi m \lambda}\tanh\left(\frac{\pi m L_y}{2}\right)W(k,m)W(n,m), \nonumber \\ \label{TVC-matrices-1} \\
%W(n,m)&=& \begin{cases} 
      %0 &  n=m, \\
      %\frac{n}{\pi}\frac{1-(-1)^n}{n^2-m^2} & n\neq m ,\\
   %\end{cases} \\
V_{nk}&=&\frac{\pi^4}{4}n^2k^2\delta_{nk}, \ \ \ \ \ \ C_{kn}=\frac{\pi^2}{4}nk\delta_{nk}, \label{TVC-matrices-2}
%C_{kn}&=&\frac{\pi^2}{4}nk\delta_{kn},
\end{eqnarray}
\end{subequations} 
where  $\delta_{nk}$ is the Kronecker delta, and $W(n,m)= \frac{n}{\pi}\frac{1-(-1)^{n+m}}{n^2-m^2}$ for $n\neq m$ and zero otherwise.

Two comments are in order regarding this Lagrangian. First, since the matrix $T_{nk}$ is coupled to the kinetic terms in the Lagrangian, it takes on the role of a mass matrix in this formulation. This mass matrix has contributions from both the inertia of the sheet, i.e., the first term in $T_{nk}$,  and the hydrodynamics of the fluid, i.e., the terms proportional to $1/\lambda$. The latter hydrodynamic terms are frequently referred to as added mass or virtual mass, since they describe an additional mass that the sheet appears to acquire when it accelerates in the fluid \cite{munk1924aerodynamic,lighthill_1960,Coene1992}. 

The second comment is related to the potential functions $\phi_i(x,y,t)$ that result from the minimization. Using Eq.~(\ref{am-sol-2}), we substitute $c_m(t)=-a_m(t)$ in the potential functions, Eq.~(\ref{phi-exp}), to obtain
\begin{equation}\label{phi-i-after-min}
\phi_i(x,y,t)=a_0(t)\left(y\pm \frac{L_y}{2}\right)+2\sum_{m=1}^{N-1}a_m(t)\cos(\pi m x)\sinh\left[\pi m\left(y\pm \frac{L_y}{2}\right)\right].
\end{equation}
This solution implies that at $y=\pm L_y/2$ the velocity of the fluid is oriented only in the $y$-direction, i.e., $\partial\phi_i/\partial x(x,\pm L_y/2,t)=0$. It also implies that $\phi_i(x,\pm L_y/2,t)=0$, which, given Eqs.~(\ref{ref-pressure}) and  (\ref{benuolli-kin-1}), yields a constant zero pressure along the inlet and the outlet walls of the chamber, $p_i(x,\pm L_y/2,t)=0$. We anticipate that these conditions will occur only when the disturbances that the sheet induces in the flow decay to zero. Therefore, the small-amplitude model holds strictly when $L_y\gg \ell$, where $\ell\simeq 1/(\pi m)$, for the smallest nonzero mode, is now explicitly identified as the decay length of the hydrodynamic disturbances. %Note that in the dimensional form $\tilde{\ell}=\tilde{L}/(\pi m)$, and so the decay length scales with the total length of the sheet.}} %Indeed, when approximatting Bernoulli's equation, we implicitly assume that the vertical dimension of the chamber  
%Therefore, while the general formulation, Eqs.~(\ref{})-(\ref{}), holds for any vertical 

% exit velocity only in the y direction
% pressure equals zero along \pm Ly/2 (c(t)=0).
% the decay length of the disturbances scales with the total length of the sheet and the mode number.
% The samll amplitude approximation holds only when Ly>>1/(\pi m).   

Keeping in mind these limitations of the small-amplitude model, we go back to derive the equations for the coefficients $A_n(t)$. Given an initial volume difference, which, in turn, corresponds to an initial configuration of the sheet, i.e., a set of initial conditions for the coefficients $A_n(0)$, 
%Given an initial configuration of the sheet, i.e., a set of initial conditions for the coefficients $A_n(0)$, 
and keeping in mind that the system starts from rest, i.e., $(dA_n/dt)(0)=0$, we can determine the dynamic evolution of the system from the minimization of Eq.~(\ref{lag-app2}) with respect to $A_n(t)$ and $F_x(t)$. This minimization yields  $N+1$ algebraic differential equations that, in our matrix notation, read:
\begin{subequations}\label{eom-An}
\begin{eqnarray}
T_{nk}\frac{d^2A_k}{dt^2}+(V_{nk}-F_x(t)C_{nk}) A_k&=&0, \label{eom-An-1} \\ \ \ \ \ \ \ \ C_{nk}A_kA_n&=&\Delta. \label{eom-An-2}
\end{eqnarray}
\end{subequations}

Once $A_n(t)$ are determined from the solution of Eq.~(\ref{eom-An}), the position of the sheet in time and in space is given by Eq.~(\ref{ysh-exp}), and the hydrodynamic potentials and the pressure fields are determined from Eqs.~(\ref{phi-exp}), (\ref{am-sol}) and (\ref{benuolli-kin-1}).  In the next section, we utilize this formulation to investigate the early time evolution. Then, in $\S$~\ref{moderate-time} we use it to analyze the dynamics at later times.

%However, before we proceed to the next section we should add a comment regarding the potential functions that are obtained after the minimization with respect to the coefficients $a_m(t)$ and $c_m(t)$. 

%add a comment regarding the limits of this expansion.  

%of the potential functions after the minimization with respect to  

\section{The early time evolution  }
\label{early-time}
In this section, we investigate the system's stability close to an initial equilibrium state. The section is divided into two parts. In the first, we recall the static solutions of the system in the small-amplitude approximation. In the second, we employ a linear stability analysis around the first two buckling modes to extract the growth rates and the flow fields of the perturbation around these modes. % and use these growth rates to obtain their corresponding flow fields.

\subsection{Recap of the quasi-static solution}
\label{recap-static}
Following the analysis in the study of Oshri~\cite{Oshri2021}, the quasi-static evolution of the system %, where $v_{\text{du}}(0)$ is the control parameter, 
is governed by two different branches of solutions, which we call ``asymmetric'' and ``symmetric''. Here, we recall the height functions in these branches.

On the one hand, when the initial volume difference is set as $0\leq v_{\text{du}}(0)\leq v_{\text{du}}^{\text{cr}}$, where $v_{\text{du}}^{\text{cr}}= \frac{2(3+\pi^2)}{\pi\sqrt{3(15+2\pi^2)}}\Delta^{1/2}$, the system is governed by the asymmetric branch. In this branch, the lateral compression is constant, $F_x(0)=4\pi^2$, and the height functions  are given by:
\begin{subequations}\label{asym-branch-static}
\begin{eqnarray}
y_{\text{sh}}(x,0)&=& \frac{p_{\text{ud}}(x,y,0)}{16\pi^4}\left[2\pi^2(1-x)x+1-\cos\left(2\pi x\right)\right]\nonumber \\ &+&\frac{1}{\pi}\sqrt{\Delta-\frac{15+2\pi^2}{768\pi^6}p_{\text{ud}}(x,y,0)^2 }\sin\left(2\pi x\right), \label{asym-branch-static-1} \\
p_{\text{ud}}(x,y,0)&=&\frac{24\pi^4}{3+\pi^2}v_{\text{du}}(0), \label{asym-branch-static-2}
\end{eqnarray}
\end{subequations}
where $p_{\text{ud}}(x,y,0)=p_{\text{u}}(x,y,0)-p_{\text{d}}(x,y,0)$ is the pressure difference between the upper and lower parts of the chamber. The potential energy of the sheet in this branch is given by:
\begin{equation}\label{energy-as-static}
E_{\text{as}}=4\pi^2\Delta-\frac{6\pi^4}{3+\pi^2}v_{\text{du}}(0)^2.
\end{equation}
Note, that when   $v_{\text{du}}(0)\rightarrow 0$ we know from Eq.~(\ref{asym-branch-static-2}) that the  pressure difference vanishes, $p_{\text{ud}}(x,y,0)\rightarrow 0$, and the elastic configuration converges to the second, asymmetric, mode of buckling, $y_{\text{sh}}(x,0)\rightarrow \sqrt{\Delta/\pi^2}\sin(2\pi x)$. The total energy of the system in this configuration is given by $E_{\text{as}}=4\pi^2\Delta$. Note also that we considered solutions with an initial  volume difference that is greater than zero. This is because the static solution has mirror symmetry around the $x$-axis. Solutions with $v_{\text{du}}(0)<0$ (and $p_{\text{ud}}(x,y,t)<0$) are  obtained by a reflection of the  height functions, Eq.~(\ref{asym-branch-static}), around the horizontal axis. 

\begin{figure}[h!]
\begin{centering}
%\captionsetup{format=plain,justification=justified}
\includegraphics[height=4.cm]{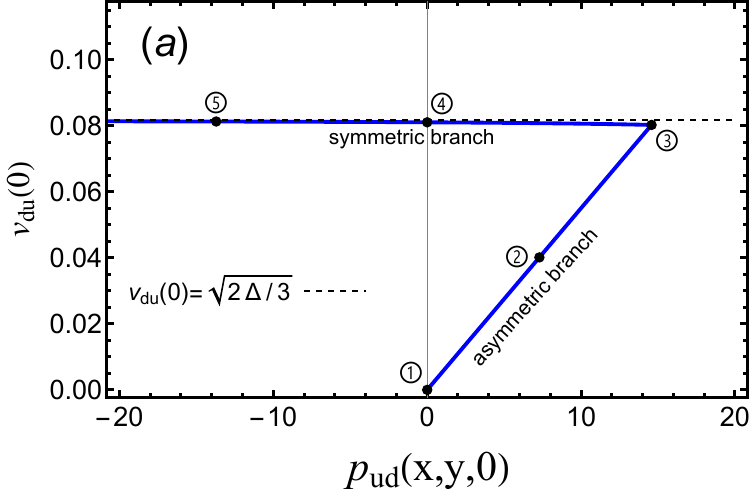} \ \ \ \ \ \ \ \ \includegraphics[height=4cm]{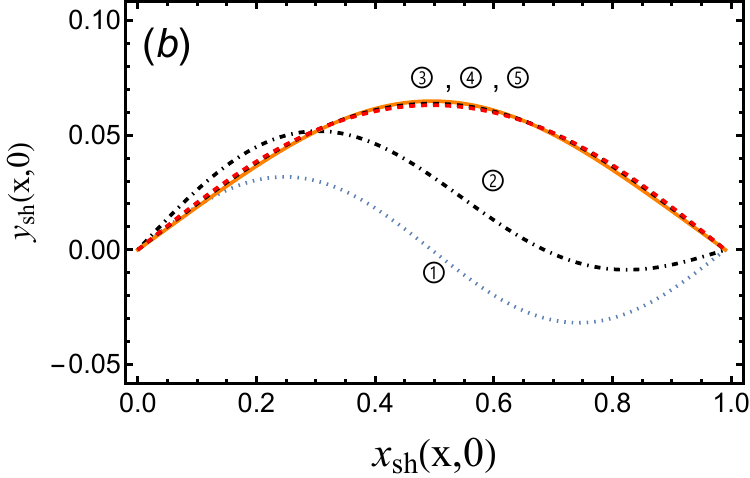}
\caption{Evolution of the static solution in the small-amplitude approximation, where $\Delta=0.01$. ({\it a}) The volume difference, $v_{\text{du}}(0)$, as a function of the pressure difference, $p_{\text{ud}}(x,y,0)$. In the asymmetric branch the $p_{\text{ud}}(x,y,0)$-$v_{\text{du}}(0)$ relation is given by Eq.~(\ref{asym-branch-static-2}), while in the symmetric branch it is given by Eqs.~(\ref{sym-branch-static-2}) and (\ref{sym-branch-static-3}). The volume difference at the asymmetric-to-symmetric transition, $v_{\text{du}}(0)=v_{\text{du}}^{\text{cr}}$, is labeled by {{\large \textcircled{\small \raisebox{0.1pt}{3}}}}. The pressure difference  in the chamber vanishes when the sheet accommodates either the second or the first mode of buckling, labels {{\large \textcircled{\small \raisebox{0.1pt}{1}}}} and {{\large \textcircled{\small \raisebox{0.1pt}{4}}}}. As the volume difference approaches its limiting value $v_{\text{du}}(0)\rightarrow (2\Delta/3)^{1/2}$, the pressure difference diverges. ({\it b}) Evolution of the sheet's profile as the volume difference increases; see the corresponding labeled numbers in panel ({\it a}). Note, that despite the relatively large change in the pressure difference in the symmetric branch, the elastic configurations remain almost  unchanged.}
\label{static-evolution}
\end{centering}
\end{figure}

On the other hand, when the volume difference is set as $v_{\text{du}}^{\text{cr}}\leq v_{\text{du}}(0)< \sqrt{2\Delta/3}$, the system is governed by the symmetric branch. In this case, the inextensibility of the sheet implies an upper limit on the volume difference. In the case of a hinged sheet, this limit is given by $\sqrt{2\Delta/3}$. The height functions in this branch are given by the parametric solution:
\begin{subequations}\label{sym-branch-static}
\begin{eqnarray}
y_{\text{sh}}(x,0)&=&\frac{p_{\text{ud}}(x,y,0)}{8u^2}(1-x)x+\frac{p_{\text{ud}}(x,y,0)}{16u^4}\left[1-\frac{\cos\left[2u(x-1/2)\right]}{\cos u}\right], \label{sym-branch-static-1}\\
p_{\text{ud}}(x,y,0)&=&-\frac{16\sqrt{6}u^{7/2}\cos u}{\sqrt{6u+4u(6+u^2)\cos^2 u-15\sin (2u)}}\Delta^{1/2},\label{sym-branch-static-2} \\
v_{\text{du}}(0)&=&-\frac{2\sqrt{2}\left[u(3+u^2)-3\tan u\right]\cos u}{\sqrt{3}u^{3/2}\sqrt{6u+4u(6+u^2)\cos^2 u-15\sin(2u)}}\Delta^{1/2}, \label{sym-branch-static-3}
\end{eqnarray}
\end{subequations}
where $u=\sqrt{F_x(0)}/2$ is a function of the lateral compression. Given the initial volume difference, $v_{\text{du}}(0)$, and the excess length, $\Delta$, we can determine the lateral compression, $F_x(0)$, from Eq.~(\ref{sym-branch-static-3}), and then substitute this solution into Eqs.~(\ref{sym-branch-static-1}) and (\ref{sym-branch-static-2}) to obtain the height profile. When $p_{\text{ud}}(x,y,0)\rightarrow 0$,  the height function converges to the first, symmetric, mode of buckling, which is given by $y_{\text{sh}}(x,0)\rightarrow \sqrt{4\Delta/\pi^2}\sin(\pi x)$ and $v_{\text{du}}(0)=8\Delta^{1/2}/\pi^2$. The total elastic energy of this shape is given by $E_{\text{s}}=\pi^2\Delta$. An example of the evolution of the sheet and the $p_{\text{ud}}(x,y,0)$-$v_{\text{du}}(0)$ relation in this static solution,  where $ 0\leq v_{\text{du}}(0)< \sqrt{2\Delta/3}$, is plotted in figure~\ref{static-evolution}.  In the following sections, we will use these height functions, Eqs.~(\ref{asym-branch-static-1}) and (\ref{sym-branch-static-1}), as the base solutions for our perturbative time-dependent expansion.

Before we proceed, we emphasize that while the sheet's configuration evolves continuously from the moment that we open the valve, the static pressure difference, given in Eqs.~(\ref{asym-branch-static-2}) and (\ref{sym-branch-static-2}), changes instantaneously at $t=0$. This is because we assumed an incompressible fluid,  in which the speed of sound is infinite. %This abrupt change in the pressures implies that static balance of forces is generally not enforced on the sheet at $t=0$. 
Nonetheless, the asymmetric and the symmetric modes of buckling, obtained respectively from Eqs.~(\ref{asym-branch-static-1}) and (\ref{sym-branch-static-1}) in the limit $p_{\text{ud}}(x,y,0)\rightarrow 0$, are exceptions. These configurations remain in static equilibrium, which can nevertheless be unstable, when the valve is opened. For this reason, in the next section, we investigate the linear stability of the system around these two limiting initial states.

\subsection{Linear stability}

To derive the linear stability around the two limiting scenarios, i.e., the second and first buckling modes, we assume that the sheet's height function is given by the static solution, up to a small perturbation that grows exponentially with time. Correspondingly, we first perturb the amplitudes of the normal modes and the lateral compression around the base solutions, i.e.,  $A_n(t)=A_n(0)+\epsilon \bar{A}_n e^{\sigma t}$ and $F_x(t)=F_x(0)+\epsilon \bar{F}_x e^{\sigma t}$, where  $\bar{A}_n$ and $\bar{F}_x$ are unknown constants, $\sigma$ is the growth rate, and $\epsilon\ll 1$ is an arbitrarily small parameter. Then, we substitute these perturbed functions into the equations of motion, Eq.~(\ref{eom-An}), and expand them up to linear order in $\epsilon$. The leading order of this expansion, order $\epsilon^0$, is given by:
\begin{subequations}\label{lin-eqn-app-epsilon0}
\begin{eqnarray}
V_{nk}A_k(0)- F_x(0)C_{nk}A_k(0)&=&0,\label{lin-eqn-app-epsilon0-1} \ \ \ \ \ \  \\  C_{nk}A_k(0)A_n(0)&=&\Delta,  \label{lin-eqn-app-epsilon0-2}
\end{eqnarray}
\end{subequations}
and the subleading order, order $\epsilon$, is given by
\begin{subequations}\label{lin-eqn-app}
\begin{eqnarray}
%\text{order $\epsilon^0$:}\ \ \ \ F_x(0)&=&\frac{V_{nk}A_k(0)}{C_{nk}A_k(0)}, \ \ \ \ \ \ C_{nk}A_k(0)A_n(0)=\Delta,  \\
\left[\sigma^2 T_{nk}+V_{nk}-F_x(0)C_{nk}\right]\bar{A}_k-C_{nk}A_k(0)\bar{F}_x&=&0,\label{lin-eqn-app-1}\\
 C_{nk}A_k(0)\bar{A}_n&=&0.\label{lin-eqn-app-2}
\end{eqnarray}
\end{subequations}
The above equations in the subleading order always have the trivial solution  $\bar{A}_n=0$ and $\bar{F}_x=0$, unless their corresponding determinant vanishes. This condition  gives  the growth rate, $\sigma$. Once $\sigma$ is determined, its corresponding eigenfunction is obtained from the solution of Eq.~(\ref{lin-eqn-app}). The hydrodynamic fields related to this eigenfunction are determined from Eqs.~(\ref{phi-exp}) and (\ref{am-sol}).

\subsubsection{Linear stability around the second mode of buckling}
\label{LS-2nd-mode}

\begin{figure}[h!]
\begin{centering}
\includegraphics[height=5cm]{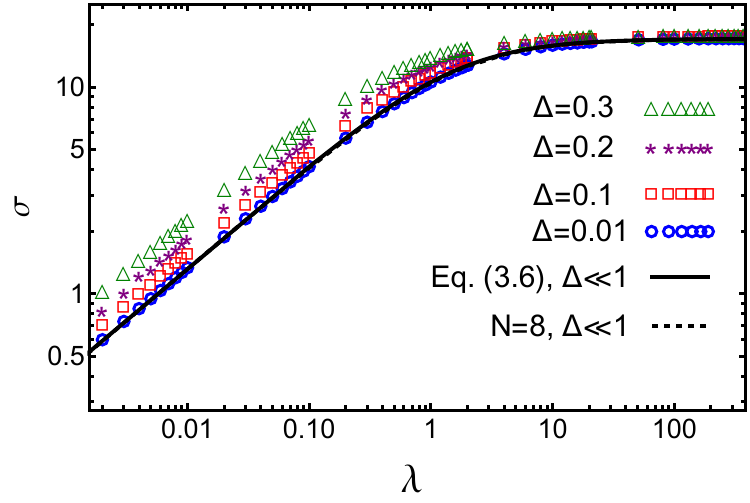} %\ \ \ \ \ \ \includegraphics[height=5cm]{eigenfunction-2nd-mode.pdf} 
%\captionsetup{format=plain,justification=justified}
\caption{Log-log plot of the highest growth rate as a function of $\lambda$ where $L_y=2$. Symbols correspond to the linear stability analysis of Eqs.~(\ref{continuity-benoulli})-(\ref{bc-sheet}), and solid and dashed lines correspond to the growth rates obtained from the small-amplitude approximation. When $\lambda\gg 1$, the growth rate converges to the constant $\sigma\simeq \sqrt{3}\pi^2$, whereas when $\lambda\ll 1$ the growth rate is given by $\sigma\simeq \sqrt{3\pi^6/8}(\lambda/L_y)^{1/2}$. Note that the differences between the solid ($N=2$) and dashed ($N=8$) lines are almost not visible in the figure. While the growth rate is independent of the excess length in the small-amplitude approximation, the more general solution (symbols), shows that the growth rate increases with $\Delta$.  }
\label{sigma-2nd-mode}
\end{centering}
\end{figure}

When the initial configuration of the sheet is given by the second mode of buckling, the base solution is derived from Eq.~(\ref{asym-branch-static-1}) in the limit $p_{\text{ud}}(x,y,0)\rightarrow 0$. This solution reads $y_{\text{sh}}(x,0)=\sqrt{\Delta/\pi^2}\sin(2\pi x)$ and $F_x(0)=4\pi^2$. A projection of this configuration on the normal mode expansion, Eq.~(\ref{ysh-exp}), gives  $A_n(0)=\sqrt{\Delta/\pi^2}\delta_{2n}$. As expected, this initial state exactly satisfies the equilibrium equations at order $\epsilon^0$, Eq.~(\ref{lin-eqn-app-epsilon0}). At the next order, i.e., order $\epsilon$, we find that  $\bar{F}_x=0$ and $\bar{A}_n=0$ for all the even perturbations, i.e., $n=2,4,6...$ . Consequently, Eq.~(\ref{lin-eqn-app}) yields linear and homogeneous equations that  involve only the odd perturbations. These equations always has the trivial solution $\bar{A}_n=0$, except when its corresponding determinant vanishes. A tractable solution to this condition, which also gives a good approximation to the highest growth rate, is obtained  at the lowest order when $N=2$. This solution reads:
\begin{equation}\label{growth-rate-2nd-app}
\sigma=\frac{\sqrt{3}\pi^2}{\sqrt{1+\frac{8 L_y}{\pi^2\lambda}}}.
\end{equation}
In figure~\ref{sigma-2nd-mode}, we plot this analytical approximation for the growth rate as a function of $\lambda$, and compare it with the numerical solution of Eq.~(\ref{lin-eqn-app}) for the case where $N=8$. In addition, we compare this analytical solution  with the growth rate obtained from the linearization of Eqs.~(\ref{continuity-benoulli})-(\ref{bc-sheet}), i.e., where $\Delta$ is assumed to be finite; see Appendix~\ref{LS-finite-Delta} for the details of this solution. 

Equation~(\ref{growth-rate-2nd-app}) is one of the central results in this paper. Several comments are in order regarding this solution. Firstly, note that the highest growth rate is always real and positive, i.e., the second mode of buckling is always an unstable state of the system. 

Secondly, while  Eq.~(\ref{growth-rate-2nd-app}) depends on the parameter $\lambda/L_y=\tilde{\rho}_{\text{sh}} \tilde{h}/(\tilde{\rho}_{\ell}\tilde{L}_y)$, this result is not general but depends on the order of approximation. Had we solved Eq.~(\ref{lin-eqn-app}) with $N\geq 3$, the two parameters $\lambda$ and $L_y$ would have appeared independently in the solution. Nonetheless, comparing the lowest order solution, i.e.,  the solution with $N=2$, with that of, say, $N=8$, we find that the growth rate remains almost unchanged (compare the solid and dashed lines in figure~\ref{sigma-2nd-mode}). For this reason, we conclude that Eq.~(\ref{growth-rate-2nd-app}) well describes the growth rate in the limit $\Delta\ll 1$. 

Thirdly, while the solution of the small-amplitude approximation is independent of the excess length, $\Delta$, the more general solution for finite values of $\Delta$ does depend on this parameter; see figure~\ref{sigma-2nd-mode} for a comparison. In particular, for a fixed value of $\lambda$, the growth rate increases with an increase in $\Delta$.

Fourthly, the analytical solution of the growth rate, Eq.~(\ref{growth-rate-2nd-app}), exhibits two different regions as a function of $\lambda$. When $\lambda/L_y\gg 1$, the growth rate is a constant, $\sigma\simeq \sqrt{3}\pi^2$, that coincides with the growth rate of a sheet that is uncoupled from an external fluid. However, when $\lambda/L_y\ll 1$, the growth rate exhibits the scaling $\sigma\simeq \sqrt{3\pi^6/8}(\lambda/L_y)^{1/2}$. These asymptotic solutions define two limiting behaviours of the system. The former scenario represents a ``solid-dominated'' region, in which the pressure difference exerted by the fluid on the sheet is negligible compared with the inertia of the sheet. The latter scenario, where $\sigma\propto (\lambda/L_y)^{1/2}$,  represents the opposite limit of a ``fluid-dominated'' region, in which the inertia of the sheet is negligible compared with the pressure difference exerted by the fluid on the sheet. Similarly, these two regions are also reflected in the added mass term, $8L_y/(\pi^2\lambda)$, in the denominator of the growth rate. When $\lambda$ is large, the added mass approaches zero and the sheet's inertia is not affected by the fluid's motion. In contrast, when $\lambda$ is small, the effective mass of the sheet increases, resulting in slower dynamics. It is worth noting that since the dynamics of the system becomes very slow in the fluid-dominated region, we would expect some aspects of this solution to align with our earlier, quasi-static solution in the asymmetric branch, as shown in Eq.~(\ref{asym-branch-static}).
%{\color{blue}{Equivalently, these two regions also manifests through the added mass $8L_y/(\pi^2\lambda)$ that appears in the denominator of the growth rate. When $\lambda$ is large the added mass diminishes to zero the sheet's inertia is unaffected by the motion of the fluid. In contrast, when $\lambda$ is small the effective mass of the sheet increases, and the dynamic is slowed down.}} Note that since the system's dynamics become very slow in the fluid-dominated region, we would expect some aspects of this solution to coincide with our quasi-static solution in the asymmetric branch, Eq. ~(\ref{asym-branch-static}). 
%This is because the quasi-static solution describes a slow spontaneous relaxation of the system, where the inertia of the sheet is negligible. This convergence to the quasi-static solution is further demonstrated below.  
This is because the quasi-static solution describes a slow spontaneous relaxation of the system, where the inertia of the sheet is negligible. This convergence to the quasi-static solution is further demonstrated in the following analysis.

\begin{figure}[h!]
\begin{centering}
\includegraphics[height=4cm]{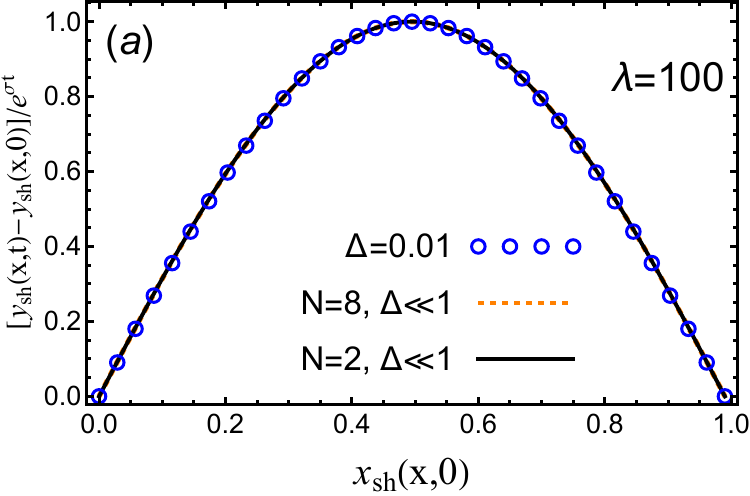}  \ \ \ \ \ \ \ \  \includegraphics[height=4cm]{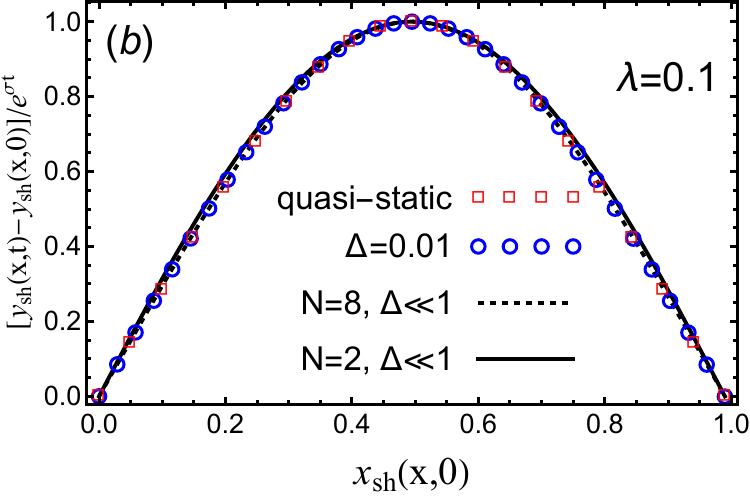} %\ \ \ 
%captionsetup{format=plain,justification=justified}
\caption{Eigenfunctions of the sheet's height function in both the solid- and fluid-dominated regions. In both panels, $L_y=2$ and open blue circles correspond to the linear stability analysis at finite $\Delta$, i.e., derived from Eqs.~(\ref{continuity-benoulli})-(\ref{bc-sheet}). The eigenfunctions are normalized such that $\left[y_{\text{sh}}(1/2,t)-y_{\text{sh}}(1/2,0)\right]/e^{\sigma t}=1$ (numerically we choose $\hat{y}_{\text{sh}}(1/2)=1$; see Appendix~\ref{LS-finite-Delta}).  ({\it a}) In the solid-dominated region $(\lambda=100)$, only one mode is excited, i.e., the low ($N=2$) and the high ($N=8$) mode approximations coincide. ({\it b}) In the fluid-dominated region ($\lambda=0.1$), all odd modes are excited. Therefore, the lowest approximation, $N=2$, does not coincide with that obtained with higher modes, $N=8$. Nonetheless, since $\bar{A}_n/\bar{A}_1\ll 1$, the differences between the low and the high orders of the  approximations are still small. Open squares represent the quasi-static approximation obtained from Eq.~(\ref{asym-branch-static}).  }
\label{eigenfunctions-2nd-mode}
\end{centering}
\end{figure}

Fifthly, note that if we fix $L_y$,  the matrices in Eqs.~(\ref{lin-eqn-app}) and (\ref{TVC-matrices}) become diagonal in the solid-dominated region. Therefore, up to small corrections of the order $1/\lambda$, $\bar{A}_1$ alone is excited at the instability. Indeed, in figure~\ref{eigenfunctions-2nd-mode}({\it a}), in which we plot the  eigenfunction for the case where $\lambda=100$ for both $N=2$ and $N=8$, we find that the two solutions are almost identical.  In contrast,  in the fluid-dominated region, $\lambda\ll 1$, and the matrices in Eq.~(\ref{lin-eqn-app}) have nonzero off-diagonal terms. Therefore, all the odd modes become coupled and are excited at the instability. Nonetheless, our numerical investigation indicates that $\bar{A}_n/\bar{A}_1\ll 1$ for all $n\geq 3$. Therefore, while we would expect the leading order solution, i.e., $N=2$, to approximate the eigenfunction well, it will not coincide with the higher-order solution. Indeed, in figure~\ref{eigenfunctions-2nd-mode}({\it b}), we compare the eigenfunctions obtained from the lowest order and the high orders of approximations and find finite differences between them. We note that the eigenfunction in the fluid-dominated region emerges from the quasi-static configuration, Eq.~(\ref{asym-branch-static-1}), when we expand Eq.~(\ref{asym-branch-static-1}) in powers of $p_{\text{ud}}(x,y,0)$, extract the linear order of this expansion, and normalize it in accordance with our convention.  The agreement between this quasi-static profile and the eigenfunction obtained from the linear stability analysis is shown in figure~\ref{eigenfunctions-2nd-mode}({\it b}) (open squares).

\begin{figure}[h!]
\begin{centering}
\includegraphics[height=7cm]{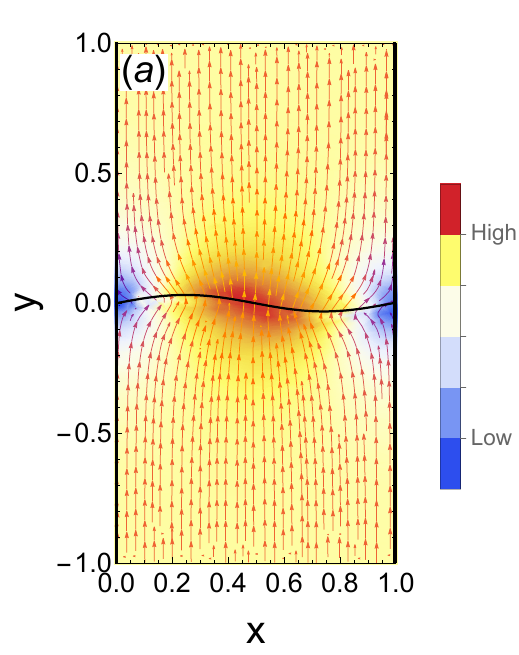}  \ \ \ \ \ \ \ \  \includegraphics[height=7cm]{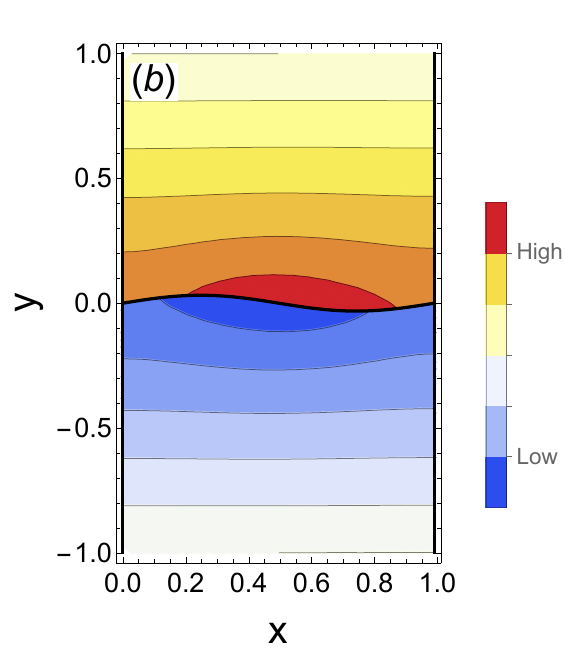} 
%\captionsetup{format=plain,justification=justified}
\caption{({\it a}) The flow field and ({\it b}) the hydrodynamic pressures obtained from the linear stability analysis of Eqs.~(\ref{continuity-benoulli})-(\ref{bc-sheet}) at the highest growth rate. In both panels, $\Delta=0.01$, $\lambda=0.1$, and $L_y=2$. The eigenfunctions are normalized as indicated in figure~\ref{eigenfunctions-2nd-mode}. This gives $\left[\text{Low, High}\right]=[0.98,2.94]$ in the color bar of the flow field, and $\left[\text{Low, High}\right]=[-78,78]$ in the color bar of the hydrodynamic pressures.  The solid black line corresponds to the initial configuration of the sheet, i.e., the asymmetric second mode of buckling. In panel ({\it a}), arrows represent the streamlines, and colors represent the relative magnitudes of the velocity. }
\label{flow-pressures-2nd-mode}
\end{centering}
\end{figure}

Sixthly, in Fig.~\ref{flow-pressures-2nd-mode}({\it a}), we plot the flow field obtained from the linear stability analysis at finite $\Delta$. Note that the maximum velocity of the fluid is obtained at the sheet's center, where the eigenfunction of the sheet's height is maximized; see figure~\ref{eigenfunctions-2nd-mode}. Unlike the growth rate, for which the small-amplitude approximation provided us with a good estimation already at $N=2$, the flow field converges at a slower pace. Convergence to the spatially dependent solution, presented in figure~\ref{flow-pressures-2nd-mode}({\it b}) is obtained only when higher modes, say $N\geq 3$, are included. This is because each  coefficient $a_m^i(t)$ in the fluid's potential functions, Eq.~(\ref{phi-exp}), depends on all the excited modes; see  Eq.~(\ref{am-sol}). In addition, in figure~\ref{flow-pressures-2nd-mode}({\it b}), we plot the eigenfunctions of the pressure fields. Note that the sheet moves upwards towards the higher pressure field. This is because the sheet's motion drives the flow, and the pressure drop in the fluid acts to slow down the onset of the elastic instability.  We note that there are no qualitative differences in the flow fields and the pressure distributions between the solid- and fluid-dominated regions. For this reason, the plots in figure~\ref{flow-pressures-2nd-mode} refer only to fluid-dominated region.

\begin{figure}[h!]
\begin{centering}
\includegraphics[height=5cm]{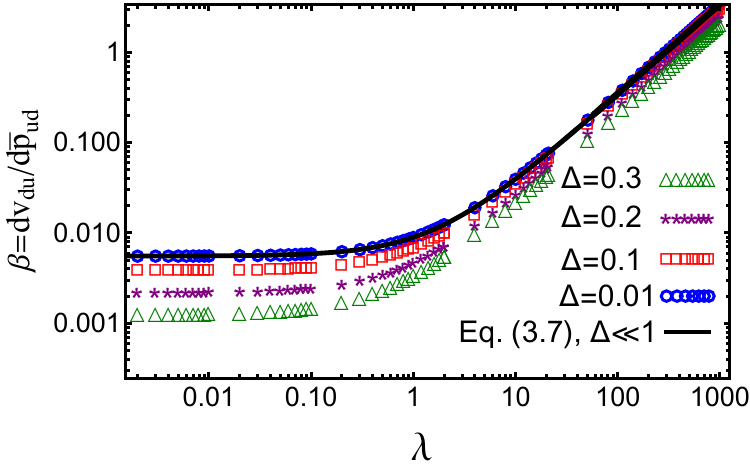}
%\captionsetup{format=plain,justification=justified}
\caption{Log-log plot of the compressibility as a function of $\lambda$ close to the onset of the instability. Symbols correspond to the compressibility at finite values of $\Delta$, and the solid black line corresponds to the analytical solution obtained from the small-slope approximation. While in the solid-dominated region $\beta\propto \lambda$, in the fluid-dominated region, $\beta$ converges to a constant.  }
\label{compressibility-2nd-mode}
\end{centering}
\end{figure}

Lastly, in addition to the growth rate, the flow field, and the hydrodynamic pressures, another experimentally measurable quantity is the system's ``compressibility'', i.e., the change in the volume difference relative to the change in the pressure difference. Since the pressure difference varies in space, we  define the compressibility as $\beta\equiv dv_{\text{du}}/d\bar{p}_{\text{ud}}$, where $\bar{p}_{\text{ud}}(t)$ is the average pressure drop on the sheet. In the small-amplitude approximation the average pressure drop on the sheet is given by $\bar{p}_{\text{ud}}(t)=\int_0^1 \left[p_{\text{u}}(x,0,t)-p_{\text{d}}(x,0,t)\right]dx$ and the volume difference in the chamber is given by $v_{\text{du}}(t)=2\int_0^1 y_{\text{sh}}(x,t)dx$ \cite{Oshri2021}. Keeping in mind that the base solution is asymmetric, and therefore does not contribute to the above integral, we find that the leading order ($N=2$) is  $v_{\text{du}}(t)/\bar{A}_1=(4/\pi)e^{\sigma t}$. In addition, the average pressure difference at this order is given by  $\bar{p}_{\text{ud}}(t)/\bar{A}_1=2L_y\sigma^2 e^{\sigma t}/(\pi\lambda)$. Consequently, at the onset of the instability,  the compressibility is a constant, which  is given by:
\begin{equation}\label{compressibility-N2} 
\beta=\frac{2\lambda}{\sigma^2 L_y}.
\end{equation}  
In figure~\ref{compressibility-2nd-mode}, we plot this result as a function of $\lambda$ and compare the lowest order approximation with the solution at finite values of $\Delta$. On the one hand, since the growth rate is constant in the solid-dominated region, we find that the compressibility scales linearly with $\lambda$, i.e.,  $\beta\rightarrow 2\lambda/(3\pi^4 L_y)$. On the other hand, since in the fluid-dominated region, the growth rate scales as $(\lambda/L_y)^{1/2}$, the compressibility converges to the constant $\beta\rightarrow 16/(3\pi^6)\simeq 5.54\times 10^{-3}$. As expected, this result for the fluid-dominated region is very close to the compressibility obtained in the quasi-static solution, Eq.~(\ref{asym-branch-static-2}), which gives $\beta=(3+\pi^2)/(24\pi^4)\simeq 5.50\times 10^{-3}$.

\subsubsection{Linear stability around the first mode of buckling}

This section analyzes the system's stability around the first mode of buckling. In contrast to the second mode, which is always unstable, the first mode represents the minimum of the elastic potential energy. Therefore, the first mode is expected to remain stable and to yield  oscillatory motion under a small dynamical perturbation. Since a purely imaginary growth rate represents this oscillatory motion,  we set  $ \sigma\equiv i\omega$, where $i=\sqrt{-1}$, and look for the lowest frequency of oscillation at the onset of the instability.

\begin{figure}[h!]
\begin{centering}
\includegraphics[height=4cm]{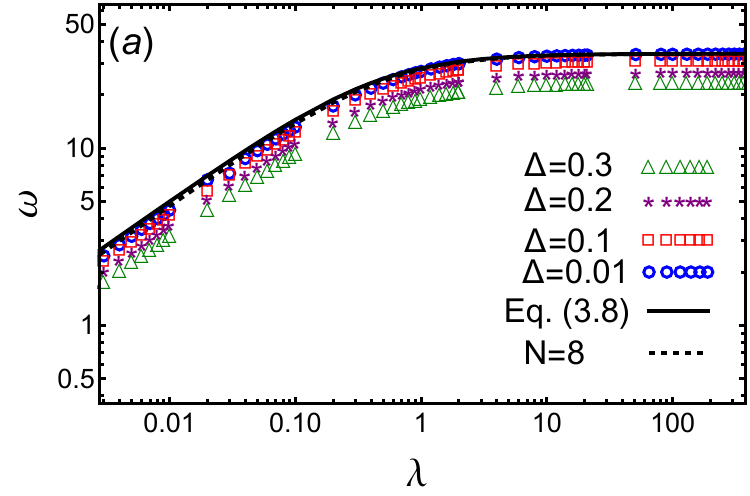} \ \ \ \ \ \ \ \ \  \includegraphics[height=4cm]{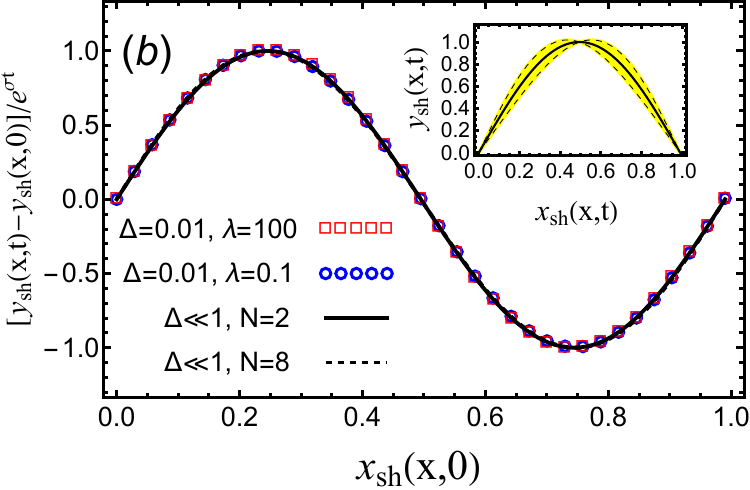} 
%\captionsetup{format=plain,justification=justified}
\caption{The lowest oscillation frequency and the sheet's eigenfunction obtained from the linear stability analysis around the first buckling mode. In both panels $L_y=2$. ({\it a}) Log-log plot of the oscillation frequency as a function of the parameter $\lambda$. Solid and dashed  lines correspond  to the solution of Eq.~(\ref{omega-1st-mode-N2}) when $N=2$ and $N=8$, respectively. All symbols correspond to the linear stability analysis of Eqs.~(\ref{continuity-benoulli})-(\ref{bc-sheet}) at finite $\Delta$. ({\it b}) The sheet's eigenfunctions in the solid- and fluid-dominated regions. All eigenfunctions are normalized such that their height at $x=1/4$ is equal to one (numerically we choose $\hat{y}_{\text{sh}}(1/4)=1$; see Appendix~\ref{LS-finite-Delta}). Although only one mode is excited when $\lambda\gg 1$ and infinite modes are excited when $\lambda \ll 1$, the eigenfunctions of the two regions are almost identical. Inset:  an example of the sheet's oscillations around the base solution. Dashed lines correspond to an illustration of the dynamic oscillations, and the solid line, to the base solution. }
\label{sigma-1st-mode}
\end{centering}
\end{figure}

When the initial state of the sheet is given by the first mode of buckling, we have from Eq.~(\ref{sym-branch-static}) that the base solution is given by $y(x,0)=(2\sqrt{\Delta}/\pi)\sin(\pi x)$ and $F_x(0)=\pi^2$. A projection of this configuration on the normal mode expansion, Eq.~(\ref{ysh-exp}), gives  $A_n(0)=(2\sqrt{\Delta}/\pi)\delta_{1n}$. This initial state, as expected, satisfies the leading order of our perturbative expansion, Eq.~(\ref{lin-eqn-app-epsilon0}). Substituting this leading order in Eq.~(\ref{lin-eqn-app}) and solving for the unknown constants, we find that $\bar{F}_x=0$ and $\bar{A}_n=0$ for all the odd modes ($n=1,3,5...$).  Consequently, Eq.~(\ref{lin-eqn-app}) yields linear and homogeneous equations that involve only the even modes of the height function. As in the previous case that we considered, a tractable solution to the vanishing determinant condition is obtained when we cut the normal mode expansion at the smallest value, $N=2$ \footnote{We note that when $L_y\gg 1$, the solution obtained from the two-mode approximation ($N=2$) is preempted by a different branch of solutions. The new branch emanates from a higher order correction in the modal expansion, and its details are beyond the scope of the present study. }. This gives,
\begin{equation}\label{omega-1st-mode-N2}
\omega= \frac{2\sqrt{3}\pi^2}{\sqrt{1+\frac{128}{9\pi^3}\frac{\tanh(\pi L_y/2)}{\lambda}}}.
\end{equation}
%and we keep in mind that the eigenvalue appears in the solution as a complex pair.  
In figure~\ref{sigma-1st-mode}({\it a}), we plot this solution and compare it with the solution of Eq.~(\ref{lin-eqn-app}) for the case of  $N=8$.  Note that our solution, Eq.~(\ref{omega-1st-mode-N2}) with $N=2$, approximates well the small-amplitude limit, $\Delta \ll 1$. Higher-order corrections, e.g., with $N=8$, almost do not alter this solution; compare the solid and dashed lines in figure~\ref{sigma-1st-mode}({\it a}). In addition, we compare Eq.~(\ref{omega-1st-mode-N2}) with the eigenvalues obtained from the linearization of Eqs.~(\ref{continuity-benoulli})-(\ref{bc-sheet}),  where the small-amplitude approximation is relaxed. We observe that as $\Delta$ increases, the oscillation frequency decreases with increasing excess length. However, the overall trend of the dependence of $\omega$ on $\lambda$ remains consistent with the small-amplitude solution. We also note that the relatively large decrease in $\omega$ when $\lambda\gg 1$ is a result of our chosen hinged boundary conditions. In contrast, systems with clamped boundary conditions display a much milder dependence on $\Delta$, as reported in previous studies \cite{NEUKIRCH2012704,Pandey_2014}.

%{\color{blue}{We find that for finite values of $\Delta$  the oscillation frequency decreases with an increase in the excess length. Nonetheless, the qualitative dependence of $\omega$ on $\lambda$ remains qualitatively similar to the small-amplitude solution. We note that the relative significant decrease in $\omega$ when $\lambda\gg 1$ is an effect of our hinged boundary conditions. A similar system, but with clamped boundary conditions, exhibits much weaker dependence on $\Delta$ \cite{NEUKIRCH2012704,Pandey_2014}. }} 

The frequency $\omega$ exemplifies the two limiting scenarios that we encountered for the growth rate in Eq.~(\ref{growth-rate-2nd-app}). On the one hand, in the solid-dominated region, where $\lambda\gg 1$, the frequency converges to the constant, $\omega \simeq 2\sqrt{3}\pi^2$, that coincides  with the frequency of oscillation of a sheet that is uncoupled from a fluid flow. On the other hand, in the fluid-dominated region, where $\lambda\ll 1$, we find the scaling, $\omega \simeq \sqrt{27\pi^7/32}\left[\lambda/\tanh(L_y\pi/2)\right]^{1/2}$, i.e., $\omega \propto \lambda^{1/2}$.  Alternatively, since the added mass is given by $\frac{128}{9\pi^3}\frac{\tanh(\pi L_y/2)}{\lambda}$, when $\lambda\gg 1$ the sheet's inertia is almost unaffected by the fluid motion, but when $\lambda\ll 1$, the added mass increases and slows down the dynamics. 
%{\color{blue}{Equivalently, since the added mass  is given by $\frac{128}{9\pi^3}\frac{\tanh(\pi L_y/2)}{\lambda}$, the inertia of the sheet is almost unaffected by the fluid when $\lambda\gg 1$. However, when $\lambda\ll 1$ the added mass increases and slows down the dynamics.}} 

These two scenarios are also manifested in the eigenfunction of the sheet's height function. In the solid-dominated region, Eq.~(\ref{lin-eqn-app}) becomes diagonal, up to corrections of the order of $1/\lambda$, and essentially only the second mode, i.e., $\bar{A}_2$, is excited at the instability, while in the fluid-dominated region Eq.~(\ref{lin-eqn-app}) has nonzero off-diagonal terms, and all the even modes are excited. Nonetheless, our numerical investigation of the solution of Eq.~(\ref{lin-eqn-app}) in the fluid-dominated region indicates that the ratio $\bar{A}_n/\bar{A}_2$ remains small for all $n$. Therefore, in both regions, deviations of the eigenfunction from the second mode are almost not visible in figure~\ref{sigma-1st-mode}({\it b}). 
 
\begin{figure}[h!]
\begin{centering}
\includegraphics[height=7cm]{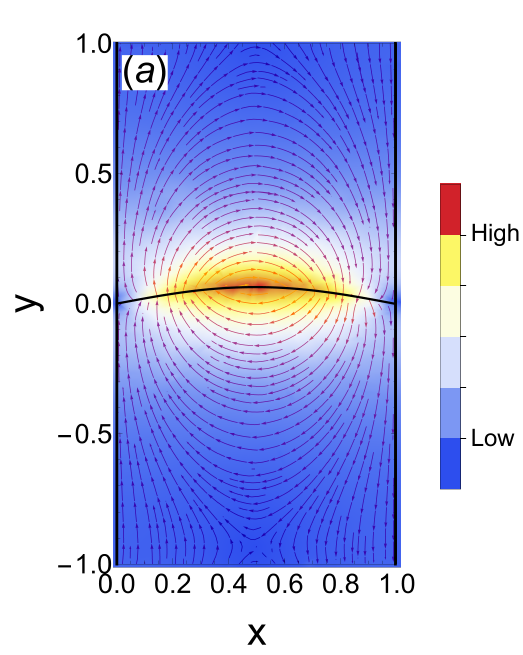} \ \ \ \ \ \ \includegraphics[height=7cm]{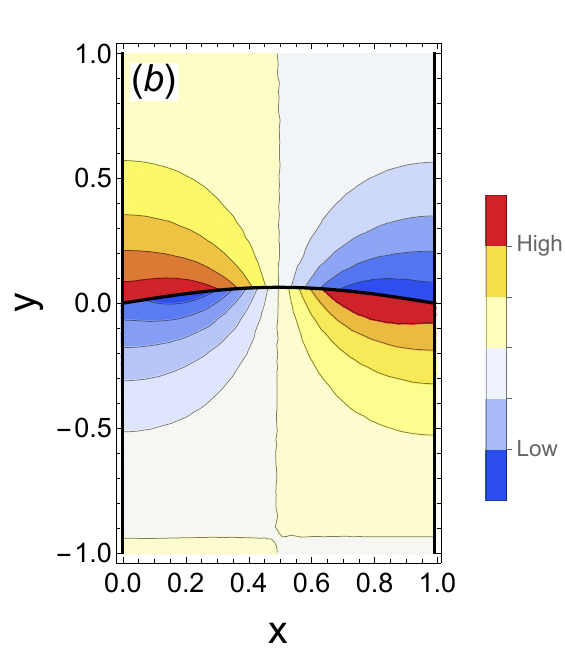} 
%\captionsetup{format=plain,justification=justified}
\caption{({\it a}) The flow fields and ({\it b}) the hydrodynamic pressure fields obtained from the linear stability analysis of Eqs.~(\ref{continuity-benoulli})-(\ref{bc-sheet}) when $\Delta=0.01$,  $\lambda=0.1$, and $L_y=2$. The normalization of the eigenfunctions is as in figure~\ref{sigma-1st-mode}({\it b}). This normalization implies $[\text{High},\text{low}]=[5.3,26.5]$ in the flow fields, and $[\text{High},\text{low}]=[-560,560]$ in the pressure fields.  In panel ({\it a}), arrows represent the streamlines, and colors represent the relative magnitudes of the velocity. }
\label{flow-pressure-1st-mode}
\end{centering}
\end{figure}

The oscillations of the sheet induce rotational flow in the chamber, whose magnitude decreases monotonically as $y$ approaches the far distant walls; see figure~\ref{flow-pressure-1st-mode}({\it a}). Indeed, the solution for $\omega$, Eq.~(\ref{sigma-1st-mode}), becomes independent of $L_y$ as the vertical dimension of the chamber increases. Furthermore, the fluid's maximum velocity is obtained close to the centre point, $x=1/2$, where the sheet's velocity equals zero. This is because the sheet moves up and down around this centre line and drives a net flux across it. The velocity of this flux is maximized at the centre of the sheet. It is important to note that there is a slight asymmetry in the flow patterns observed in the upper and lower regions of the chamber, which may be attributed to the initial non-zero volume difference in the base solution. 
%{\color{blue}{Note that there is a slight asymmetry between the flow fields in the upper and lower parts of the chamber that results from the nonzero volume difference in the base solution. %This asymmetry is absent from the flow fields obtained in the small-amplitude approximation, Eq.~(\ref{phi-i-after-min}), which satisfy $\phi_{\text{d}}(x,-y,t)=-\phi_{\text{u}}(x,y,t)$.
%}}

%originate from a higher order correction compared with the small-amplitude approximation, Eq.~(\ref{phi-i-after-min}). In this approximation, the flow fields above and below the sheet satisfy the symmetry $\phi_{\text{d}}(x,-y,t)=-\phi_{\text{u}}(x,y,t)$.  }}

The pressure fields induced by the elastic oscillations are plotted in figure~\ref{flow-pressure-1st-mode}({\it b}) and show an asymmetric profile in correlation with the eigenfunction of the sheet (figure~\ref{sigma-1st-mode}({\it b})). Since only even modes are excited at the instability, the average pressure difference on the sheet, $\bar{p}_{\text{ud}}(t)$, vanishes. Therefore, in this case, there is no analogue to the compressibility calculated in $\S$~\ref{LS-2nd-mode}.

\section{The evolution at moderate times }
\label{moderate-time}

In this section, we relax the assumption that $t\ll 1$ and extend the analysis up to moderate times. The limits of this analysis are discussed at the end of this section. %$\sigma t\sim O(1)$, where $\sigma$ is the growth rate of the instability. The exact limits of this moderate time analysis are discussed at the end of this section. 
 In particular, we require that the initial configuration of the sheet be close in shape to the second mode of buckling, i.e., $v_{\text{du}}(0)\ll v_{\text{du}}^{\text{cr}}$, where $v_{\text{du}}^{\text{cr}}=\frac{2(3+\pi^2)}{\pi\sqrt{3(15+2\pi^2)}}\Delta^{1/2}$ (see $\S$~\ref{recap-static}), and we investigate the following questions: (i) What is the maximum amount of energy that is transferred from the sheet to the fluid? (ii) How long does it take the system to convert this maximum elastic energy into a fluid flow? (iii) What is the time-dependent behaviour of the $\bar{p}_{\text{ud}}(t)$-$v_{\text{du}}(t)$ relation. To address these questions,  we assume that the amplitude of the sheet remains small during the dynamic evolution of the system and utilize the approximated formulation derived in $\S$~\ref{small-slope-formulation}. This formulation yields the simplified set of nonlinear equations, Eq.~(\ref{eom-An}), that describe the coupling between the elastic and the hydrodynamic equations. As may be seen, this set of equations has a conserved first integral, $E=T_{nk}\frac{dA_k}{dt}\frac{dA_n}{dt}+V_{nk}A_k A_n$, that corresponds to the total energy of the system, Eq.~(\ref{energy-tot}).  This conserved energy constitutes the starting point for the discussion that follows.

When the initial configuration of the sheet is close in shape to the second mode of buckling, we expect the system's dynamics at moderate times to depend strongly on the first two modes, $A_1(t)$ and $A_2(t)$. This is because the initial shape of the sheet and its corresponding eigenfunction at the highest growth rate are described approximately by these two modes; see figure~\ref{eigenfunctions-2nd-mode}. Therefore, we reduce the expression for the total energy to the case in which $N=2$ and obtain the following equation:
\begin{equation}\label{two-modes-approx}
E=\frac{1}{4}\left(1+\frac{8L_y}{\pi^2 \lambda}\right)\left(\frac{dA_1}{dt}\right)^2+\frac{1}{4}\left(1+\frac{128\tanh(\pi L_y/2)}{9\pi^3\lambda}\right)\left(\frac{dA_2}{dt}\right)^2+\frac{\pi^4}{4}A_1^2+4\pi^4A_2^2.
%E-E_{\min}=\left[\frac{1}{4}+\frac{32\tanh(\pi L_y/2)}{9\pi^3\lambda}+\frac{(8L_y+\pi^2\lambda)A_2^2}{\lambda(\Delta-\pi^2A_2^2)}\right]\left(\frac{dA_2}{dt}\right)^2+3\pi^4 A_2^2,
\end{equation} 
We keep in mind that $A_1(t)$ and $A_2(t)$ are related through the constraint of the excess length, Eq.~(\ref{eom-An-2}).

\begin{figure}[h!]
\begin{centering}
\includegraphics[height=4cm]{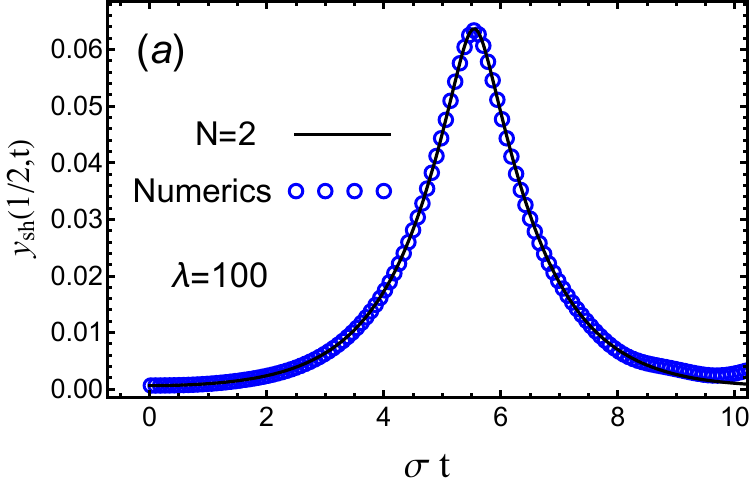} \ \ \ \ \ \ \includegraphics[height=4cm]{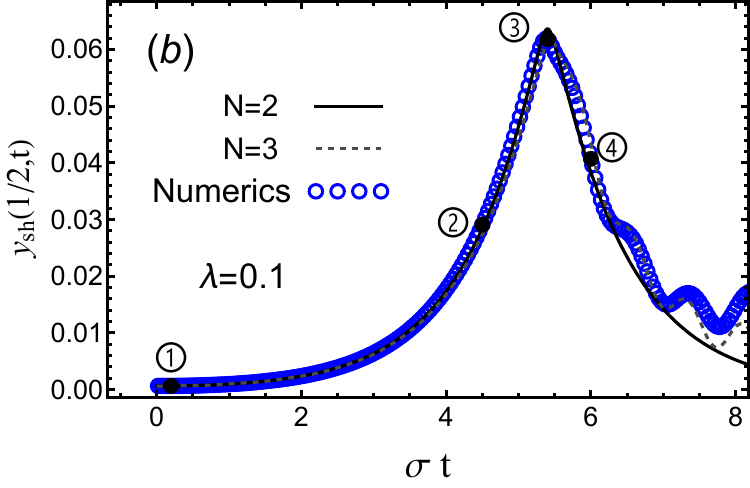} \ \ \ \ \ \ \includegraphics[height=4cm]{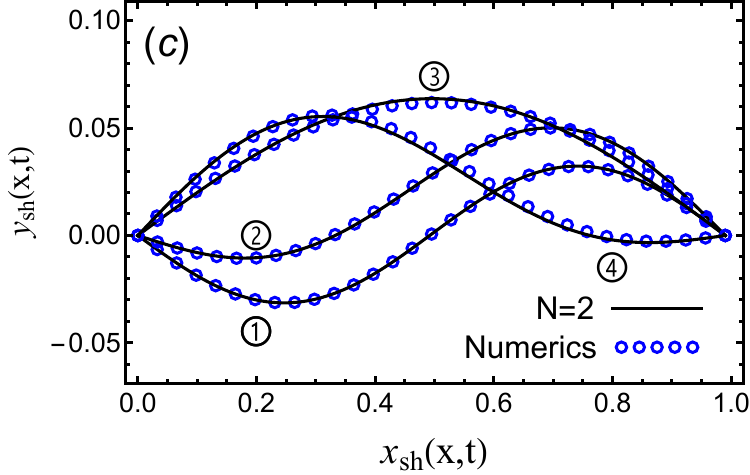} 
%\captionsetup{format=plain,justification=justified}
\caption{The sheet's mid-point as a function of time in ({\it a}) the solid-dominated ($\lambda=100$) and ({\it b}) the fluid-dominated ($\lambda=0.1$) regions. In both panels, $\Delta=0.01$, $L_y=2$, and  the growth rate, $\sigma$, is approximated by Eq.~(\ref{growth-rate-2nd-app}). The solid black line denotes the two-mode approximation, i.e.,  Eq.~(\ref{eom-An}) with $N=2$,  and the open blue circles denote the solid black line denotes the solution of the nonlinear model, Eqs.~(\ref{continuity-benoulli})-(\ref{bc-sheet}). In panel (b), the dashed gray line denotes the solution of Eq.~(\ref{eom-An}) with $N=3$. The initial configuration of the sheet is given by Eq.~(\ref{asym-branch-static}) with $v_{\text{du}}(0)=0.01v_{\text{du}}^{\text{cr}}$. In both cases, the approximated solution breaks down after $y_{\text{sh}}(1/2,t)$ reaches its first maximum. In the solid-dominated region the two-mode approximation holds for longer times compared with the fluid-dominated region.  (c) The configurations of the sheet along the trajectory depicted in panel (b); see the corresponding markers in panel (b). Between {{\large \textcircled{\small \raisebox{0.1pt}{1}}}}-{{\large \textcircled{\small \raisebox{0.1pt}{3}}}} the sheet releases potential energy as it transforms from the second mode of buckling to the first mode of buckling. After {{\large \textcircled{\small \raisebox{0.1pt}{3}}}}, the height of the sheet's mid-point decreases and the sheet gains back potential energy, as seen in  {{\large \textcircled{\small \raisebox{0.1pt}{4}}}}.
   }
\label{ysh-half-two-lambda}
\end{centering}
\end{figure}

To obtain some insight regarding the validity of this two-mode approximation, we solve  Eq.~(\ref{eom-An}) numerically with $N=2$ and compare the results with the solution of the nonlinear model, i.e., the numerical solution of Eqs.~(\ref{continuity-benoulli})-(\ref{bc-sheet}). In our investigation,  we set $\Delta=0.01$ and $L_y=2$, and consider two different values for the parameter $\lambda$.  The initial configuration is given by Eq.~(\ref{asym-branch-static}), where $v_{\text{du}}(0)=0.01v_{\text{du}}^{\text{cr}}$ \footnote{The initial conditions are given by $A_1(0)=2\int_0^1 y_{\text{sh}}(x,0)\sin(\pi x)dx=\frac{32v_{\text{du}}(0)}{3\pi+\pi^3}$ and $A_2(0)=\sqrt{\frac{\Delta}{\pi^2}-\frac{256v_{\text{du}}(0)^2}{(3\pi+\pi^3)^2}}$, such that Eq.~(\ref{eom-An-2}) is satisfied. In addition, we keep in mind that the system starts from rest, $\frac{dA_1}{dt}(0)=\frac{dA_2}{dt}(0)=0$.}. The results of these numerical solutions are presented in figures~\ref{ysh-half-two-lambda}(a) and~\ref{ysh-half-two-lambda}(b), where we follow the time-dependent behaviour of the mid-point on the sheet, $y_{\text{sh}}(1/2,t)$. The configurations of the sheet along the trajectory depicted in figure~\ref{ysh-half-two-lambda}(b) are presented in figure~\ref{ysh-half-two-lambda}(c). In both cases, we find that the approximated solution breaks down slightly after $y_{\text{sh}}(1/2,t)$ reaches its first maximum. In the solid-dominated region ($\lambda=100$), the two-mode approximation holds over one period of motion, while in the fluid-dominated region ($\lambda=0.1$), the approximation breaks down a little earlier. This difference is probably due to the different number of excited modes at the instability in each region of the system; see discussion in $\S$~\ref{LS-2nd-mode}. %While in the fluid-dominated region infinite number of modes are excited at the instability, in the solid-dominated region, only two modes are essentially excited. 
We conjecture that higher modes become active beyond moderate times in the fluid-dominated region and perturb the system's trajectory from the two-mode approximation. Indeed, by solving Eq.(\ref{eom-An}) with $N=3$ instead of $N=2$, we observe a convergence towards the numerical data for longer times; see the dashed-gray line in figure~\ref{ysh-half-two-lambda}(b).

Nonetheless, in both cases, i.e., the solid- and fluid-dominated regions, the agreement between the numerical solution of Eqs.~(\ref{continuity-benoulli})-(\ref{bc-sheet}) and the analytical approximation, Eq.~(\ref{eom-An}) with $N=2$, holds up to the first maximum. Similar  results are also obtained when we perturb the system's parameters, $\Delta$ and $L_y$, and the initial configuration. Therefore, in the following analysis, we will utilize this approximation to examine the system's behaviour up to the point where the midpoint of the sheet reaches its first maximum. We refer to this stage of the system as moderate times.  
%dynamics at moderate times while keeping its limitations in mind.

\subsection{The elasto-hydrodynamic energetic interplay}
\label{elasto-hydrodynamic-energetic-interplay}
Since the initial configuration of the sheet is close in shape to the second mode of buckling and the total energy of the system is conserved, the total energy at any $t>0$ is given by Eq.~(\ref{energy-as-static}). In the limit  $v_{\text{du}}(0)\ll v_{\text{du}}^{\text{cr}}$,  this energy is approximated as $E_{\text{as}}\simeq 4\pi^2 \Delta$, i.e., the energy of the second mode of buckling.  In addition, since the first mode of buckling is the global minimizer of the elastic sheet's potential energy, the potential energy of the sheet cannot fall below  $E_{\text{s}}=\pi^2\Delta$. Therefore, at most, our system can convert $\delta E=E_{\text{as}}-E_{\text{s}}\simeq 3\pi^2 \Delta$ of the initial potential energy either into kinetic energy of the sheet or into hydrodynamic energy of the fluid. We remind the reader that the kinetic and potential energies of the sheet, $E_{\text{sh}}^{\text{k}}(t)$ and $E_{\text{sh}}^{\text{p}}(t)$, and the energy of the fluid, $E_{\text{f}}(t)$,  are given  by the first, second, and third terms, respectively, in the right-hand side of Eq.~(\ref{energy-tot}).  In the small-amplitude approximation, these energies reduce to $E_{\text{sh}}^{\text{k}}(t)=\left(\lim_{\lambda\rightarrow \infty} T_{kn}\right)\frac{dA_k}{dt}\frac{dA_n}{dt}$, $E_{\text{sh}}^{\text{p}}(t)=V_{kn}A_k A_n$, and $E_{\text{f}}(t)=T_{kn}\frac{dA_k}{dt}\frac{dA_n}{dt}-E_{\text{sh}}^{\text{k}}$. 

\begin{figure}[h!]
\begin{centering}
\includegraphics[height=3.75cm]{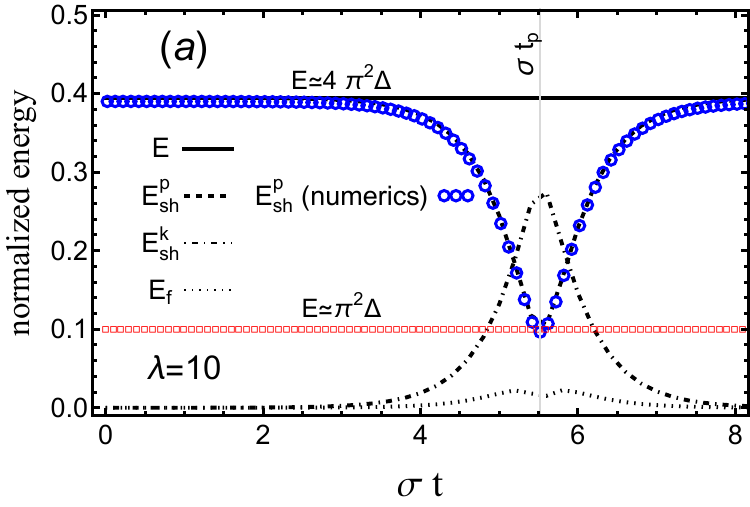} \ \ \ \ \ \ \includegraphics[height=3.75cm]{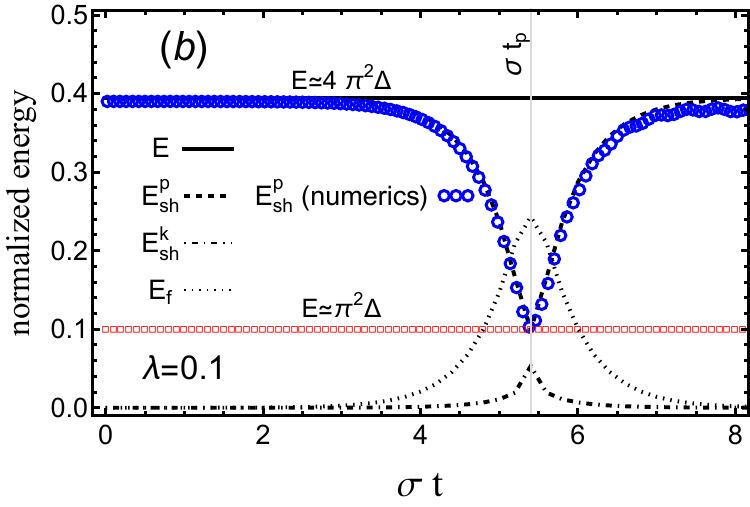}\ \ \ \ \ \ \includegraphics[height=4.05cm]{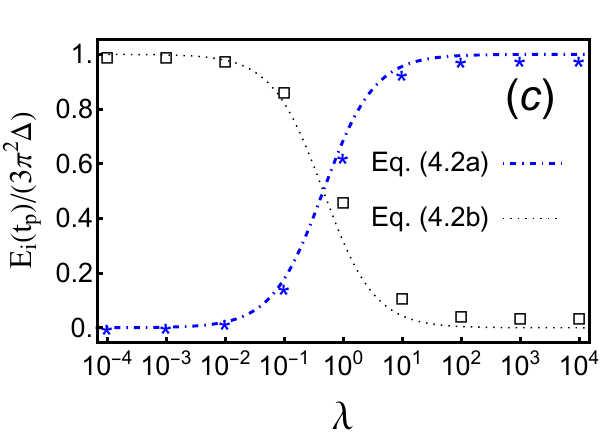}  
%\captionsetup{format=plain,justification=justified}
\caption{The energetic interplay between the three components of the total energy in ({\it a}) the solid-dominated ($\lambda=10$) and ({\it b}) the fluid-dominated ($\lambda=0.1$) regions of the system. In these two panels, we solve Eq.~(\ref{eom-An}) for the case where $N=2$, $\Delta=0.01$, $L_y=2$, and $\sigma$ is given by Eq.~(\ref{growth-rate-2nd-app}). The initial configuration of the sheet is given by Eq.~(\ref{asym-branch-static}), where $v_{\text{du}}(0)=0.01v_{\text{du}}^{\text{cr}}$. After some initial delay, the potential energy of the sheet drops from $E_{\text{sh}}^{\text{p}}(t\ll 1)\simeq 4\pi^2 \Delta$ to $E_{\text{sh}}^{\text{p}}( \text{t}_p)\simeq \pi^2\Delta$. Open blue circles represent the potential energy obtained from the solution of the nonlinear equations, Eqs.~(\ref{continuity-benoulli})-(\ref{bc-sheet}). The energy released from the sheet is divided between the kinetic energy of the sheet, $E_{\text{sh}}^{\text{k}}(t)$, and the hydrodynamic energy of the fluid, $E_{\text{f}}(t)$, such that the total energy, $E$, remains constant. In the solid-dominated region, $E_{\text{sh}}^{\text{k}}(t_{\text{p}})\gg E_{\text{f}}(t_{\text{p}})$, while in the fluid-dominated region we find the opposite relation, $E_{\text{sh}}^{\text{k}}(t_{\text{p}})\ll E_{\text{f}}(t_{\text{p}})$. ({\it c}) The kinetic energy of the sheet and the fluid at $t=t_{\text{p}}$ as a function of $\lambda$, i.e., $E_i(t_{\text{p}})=E_{\text{sh}}^{\text{k}}(t_{\text{p}}), E_{\text{f}}(t_{\text{p}})$. A logarithmic scale is used on the $x$-axis. The dotted and the dashed-dotted lines correspond to our analytical solution from the two-mode approximation, and the color symbols correspond to the numerical solution of Eqs.~(\ref{continuity-benoulli})-(\ref{bc-sheet}), where the initial conditions are similar to those used in panels ({\it a}) and ({\it b}).  }
\label{energies}
\end{centering}
\end{figure}

The typical evolution of the three components of the energy, i.e., the kinetic energy of the sheet, the potential energy of the sheet, and the energy of the fluid, is plotted in figures~\ref{energies}({\it a}) and \ref{energies}({\it b}). These plots are obtained from the solution of Eq.~(\ref{eom-An}) with $N=2$ in the  solid- and the fluid-dominated regions,  $\lambda=10$ and $\lambda=0.1$ respectively, where the initial configuration is given by Eq.~(\ref{asym-branch-static}) with $v_{\text{du}}(0)=0.01v_{\text{du}}^{\text{cr}}$. In both cases, we find that, after some initial delay, the potential energy of the sheet drops from $E\simeq 4\pi^2 \Delta$, i.e., the energy of the second mode of buckling, Eq.~(\ref{energy-as-static}), to $E\simeq \pi^2\Delta$, i.e., the energy of the first mode of buckling. The energy released in this process, $\delta E\simeq 3\pi^2\Delta$, is converted to the kinetic energy of the sheet and the energy of the fluid, while the total energy remains fixed. In the solid-dominated region (figure~\ref{energies}({\it a})) the kinetic energy of the sheet becomes much larger than the energy of the fluid, while as $\lambda$ decreases,  the opposite picture emerges, i.e., the energy of the fluid becomes much larger than the kinetic energy of the sheet  (figure~\ref{energies}({\it b})). In both scenarios, shortly after the initial peak, a portion of the kinetic energy is converted back into potential energy of the sheet. As a result, the sheet tends to return to its configuration of the second buckling mode, thereby decreasing the height of the sheet's midpoint, as shown in figure~\ref{ysh-half-two-lambda}(c).

%{\color{blue}{In both cases, slightly after the first peak, the kinetic energies are converted back into the potential energy of the sheet. In return, the sheet tends to recover the configuration of the second buckling mode, thereby reducing the height of the sheet's mid-point, see Fig.~\ref{ysh-half-two-lambda}.}}

Our two-mode approximation allows us to quantify these findings and to further estimate the maximum values of $E_{\text{sh}}^{\text{k}}( t_{\text{p}})$ and $E_{\text{f}}(t_{\text{p}})$ as a function of $\lambda$, where  $t_{\text{p}}$ denotes the time at which the  potential energy of the sheet reaches its minimum value. Indeed, at $E_{\text{sh}}^{\text{p}}(t_{\text{p}})=\pi^2\Delta$, the sheet is close in shape to the first mode of buckling, i.e., $A_1(t_{\text{p}})\simeq 2\sqrt{\Delta}/\pi$. The constraint on the excess length, Eq.~(\ref{eom-An-2}), then implies that $A_2(t_{\text{p}})\simeq 0$ and that $(dA_1/dt)(t_{\text{p}})\simeq 0$. In addition, the derivative of the second mode at that moment, $(dA_2/dt)(t_{\text{p}})$, is obtained from  Eq.~(\ref{two-modes-approx}) when we substitute $E=4\pi^2\Delta$ for the total energy. Taken together, these approximations yield the following maximum energies at $t=t_{\text{p}}$:
\begin{subequations}\label{energies-Ek-Ef}
\begin{eqnarray}
\frac{E_{\text{sh}}^{\text{k}}(t_{\text{p}})}{3\pi^2\Delta}&=&\frac{1}{1+\frac{128 \tanh(\pi L_y/2)}{9\pi^3\lambda}}, \label{energies-Ek-Ef-1}\\
\frac{E_{\text{f}}(t_{\text{p}})}{3\pi^2\Delta}&=&\frac{1}{1+\frac{9\pi^3\lambda}{128 \tanh(\pi L_y/2)}}. \label{energies-Ek-Ef-2}
\end{eqnarray}
\end{subequations}
In figure~\ref{energies}({\it c}), we compare these maximum energies with the numerical solution of Eqs.~(\ref{continuity-benoulli})-(\ref{bc-sheet}). Overall, we find a good fit between the analytical  approximation and the numerical solution over the entire range of $\lambda$.

Using Eq.~(\ref{energies-Ek-Ef}), we find that in the solid-dominated region, $\lambda\gg 1$, most of the initial energy is converted into the kinetic energy of the sheet, i.e., $E_{\text{sh}}^{\text{k}}(t_{\text{p}})\simeq 3\pi^2 \Delta -\delta_1 $ and $E_{\text{f}}(t_{\text{p}})\simeq \delta_1$, where $\delta_1=128\Delta\tanh(\pi L_y/2)/(3\pi\lambda)$, while in the fluid-dominated region, $\lambda\ll 1$, most of the energy is converted into the energy of the fluid, i.e., $E_{\text{sh}}^{\text{k}}(t_{\text{p}})\simeq \delta_2$ and $E_{\text{f}}(t_{\text{p}})\simeq 3\pi^2\Delta-\delta_2$, where $\delta_2=27\pi^5\Delta\lambda/\left[128\tanh(\pi L_y/2)\right]$. Furthermore, if we estimate the average instantaneous velocity of the fluid, $\bar{v}(t_{\text{p}})$, by using $ E_{\text{f}}(t_{\text{p}})\propto \bar{v}(t_{\text{p}})^2 \ell/\lambda$,  we find  that in the solid-dominated region the average velocity is proportional to $\bar{v}(t_{\text{p}})\propto(\Delta/\ell)^{1/2} $, while in the fluid-dominated region it is proportional to $\bar{v}(t_{\text{p}})\propto (\lambda \Delta/\ell)^{1/2}$; namely, while $E_{\text{f}}(t_{\text{p}},\lambda\ll 1)\gg E_{\text{f}}(t_{\text{p}},\lambda\gg 1)$,  their corresponding velocities present the opposite relationship, i.e., $\bar{v}(t_{\text{p}},\lambda\ll 1)\ll \bar{v}(t_{\text{p}},\lambda\gg 1)$. This is because the momentum of the fluid, $p_{\text{f}}\propto\bar{v}(t_{\text{p}})/\lambda$ - but not its velocity - increases in the fluid-dominated region.

\subsection{The peak time}

%{\color{blue}{While in the previous section, we showed, using energetic considerations, that the sheet tends to release almost all its available potential energy, in this section, we ask how long it takes the system to release this stored energy. }} 
In the previous section, we demonstrated through energetic considerations that the sheet tends to release most of its stored potential energy. In this section, we investigate the time it takes for the system to release this energy. Given an initial configuration of the sheet that is close in shape to the second mode of buckling, i.e., Eq.~(\ref{asym-branch-static}) with $v_{\text{du}}(0)\ll v_{\text{du}}^{\text{cr}}$, we aim to find the time $t=t_{\text{p}}$ at which the potential energy of the sheet first drops to the minimum value, $E_{\text{sh}}^{\text{p}}(t_{\text{p}})\simeq \pi^2 \Delta$. To do so, we  first use Eqs.~(\ref{eom-An-2})  and (\ref{growth-rate-2nd-app}) to  eliminate   $A_2(t)$ and $\lambda$ in favor of $A_1(t)$ and $\sigma$, respectively, in Eq.~(\ref{two-modes-approx}). Then, we substitute the energy of the initial configuration, Eq.~(\ref{energy-as-static}), into  Eq.~(\ref{two-modes-approx}) and integrate it between $t\in[0,t_{\text{p}}]$. This gives:
\begin{equation}\label{sigma-tp-analytic}
\sigma t_{\text{p}}=\int_{A_1(0)}^{A_1(t_\text{p})}\frac{\sqrt{432\pi^3\Delta+\left[9\pi\left(-12\pi^4+\sigma^2\right)+\frac{16\left(3\pi^4-\sigma^2\right)}{L_y\coth(\pi L_y/2)}\right]A_1^2}}{6\sqrt{3}\pi^{3/2}\sqrt{\left(A_1^2-\frac{8v_{\text{du}}(0)^2}{3+\pi^2}\right)\left(4\Delta-\pi^2A_1^2\right)}}dA_1,
\end{equation}
where $A_1(t_{\text{p}})\simeq 2\Delta^{1/2}/\pi$ is approximately the  amplitude of the sheet at time $t_{\text{p}}$, and $A_1(0)=2\int_0^1 y_{\text{sh}}(x,0)\sin(\pi x)dx=32v_{\text{du}}(0)/(3\pi+\pi^3)$ is the projection of  the initial configuration, Eq.~(\ref{asym-branch-static}), on the first mode of the sheet.

\begin{figure}[h!]
\begin{centering}
\includegraphics[height=4cm]{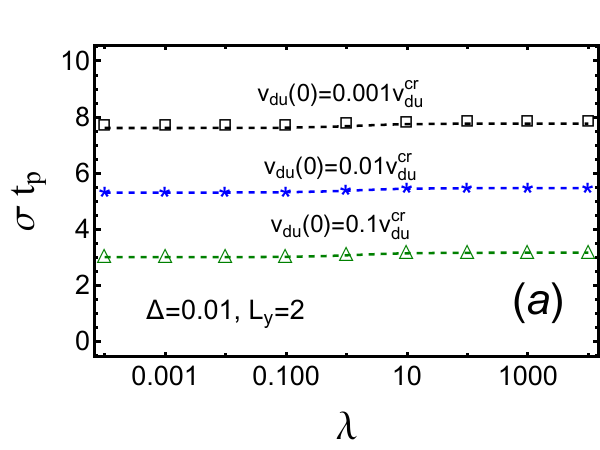} \ \ \  \ \ \ \ \ \includegraphics[height=4cm]{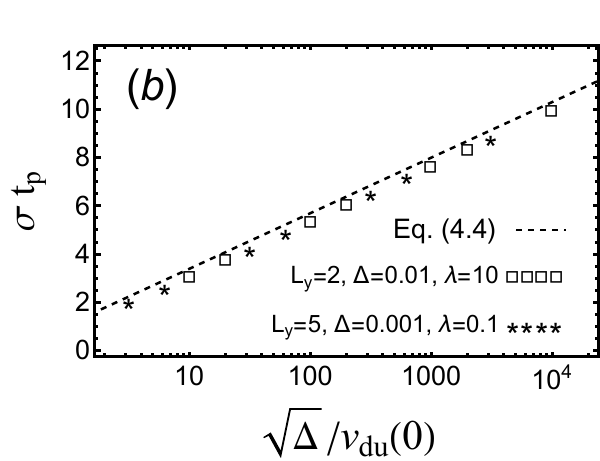} 
%\captionsetup{format=plain,justification=justified}
\caption{The peak time,  $\sigma t_{\text{p}}$, as a function of the system's parameters. ({\it a}) The peak time as a function of $\lambda$, where $\Delta=0.01$ and $L_y=2$, for two different values of the initial volume difference. Dashed lines correspond to the numerical integration of Eq.~(\ref{sigma-tp-analytic}), and symbols with corresponding colors represent the numerical solution of Eqs.~(\ref{continuity-benoulli})-(\ref{bc-sheet}). For a given $v_{\text{du}}(0)$, the time $\sigma t_{\text{p}}$ changes by less than $5\% $ over six orders of magnitude of the parameter $\lambda$. ({\it b}) Comparison between the analytical scaling, Eq.~(\ref{scaling-sigma-tp}), and the solution obtained from the nonlinear model. A logarithmic scale is used on the $x$-axis. Symbols correspond to the numerical solution of the nonlinear model. While the scaling of the analytical solution agrees well with the numerical data, the prefactor $c\simeq 2.9$ slightly overestimates the numerical prediction.   }
\label{tp-vs-lambda}
\end{centering}
\end{figure}

In figure~\ref{tp-vs-lambda}({\it a}), we fix the excess length and the vertical dimension of the chamber at $\Delta=0.01$ and $L_y=2$, respectively, and plot $\sigma t_{\text{p}}$ for $\lambda\in [10^{-3},10^3]$ ($\sigma\in[0.4,\sqrt{3}\pi^2]$). We find that at a given initial volume difference, $v_{\text{du}}(0)$,  the peak time, $\sigma t_{\text{p}}$, changes by less than five percent over more than six orders of magnitude in $\lambda$. While $\sigma t_{\text{p}}$ is almost independent of $\lambda$, it does  depend strongly on the initial configuration of the sheet, $v_{\text{du}}(0)$, and the excess length, $\Delta$. An analytical approximation of this dependence can be extracted from Eq.~(\ref{sigma-tp-analytic}), if we assume that the system is in the solid-dominated region, where $\sigma \simeq \sqrt{3}\pi^2$. Under this assumption, we can integrate the right-hand side of this equation  and take the limit $v_{\text{du}}(0)\ll 1$ of the resulting expression \footnote{We use {\it Mathematica} \cite{Mathematica} for the symbolic integration. }. This gives:
\begin{equation}\label{scaling-sigma-tp}
\sigma t_{\text{p}}\simeq   \ln \left(c \frac{\Delta^{1/2}}{v_{\text{du}}(0)}\right),
\end{equation}
where $c\simeq 2.9$. While the scaling in Eq.~(\ref{scaling-sigma-tp}) agrees well with our numerical solution of the nonlinear model, there is a small deviation in the numerical prefactor. The best fit to the nonlinear model gives $c\simeq 2.0$; see figure~\ref{tp-vs-lambda}(b). We note that the independence of $L_y$ in the right-hand side of Eq.~(\ref{scaling-sigma-tp}) is a result of our assumption that $\lambda\gg 1$. Had we derived this scaling using the fluid-dominated region, we would have found that the integral in Eq.~(\ref{sigma-tp-analytic}) does depend on the vertical dimension of the chamber. However, this dependence diminishes to zero when $L_y\gtrsim 1$, as is required by our fourth assumption in \S~\ref{formulation}.

We also note that the logarithmic divergence of $\sigma t_{\text{p}}$ in the limit $v_{\text{du}}(0)\rightarrow 0$ is similar to the divergence of the period of a pendulum near the separatrix \cite{Butikov_1999}. Within this analogy between the two problems, the small initial deviation of the pendulum from the unstable vertical position is analogous to the small initial volume difference $v_{\text{du}}(0)$. The separatrix of the pendulum is analogous to the trajectory $v_{\text{du}}(0)=0$ in the phase space of our system.

\subsection{The $\bar{p}_{\text{ud}}$-$v_{\text{du}}$ relation}

In this section, we investigate the behaviour of the  $\bar{p}_{\text{ud}}$-$v_{\text{du}}$ relation at moderate times. To this end, we first use the two-mode approximation to calculate the pressure difference in the chamber. From Bernoulli's  equation, Eq.~(\ref{benuolli-kin-1}), and the normal mode expansion, Eqs.~(\ref{phi-exp}) and (\ref{am-sol}), we have that  the average pressure difference on the sheet is given by $\bar{p}_{\text{ud}}(t)=\frac{2L_y}{\pi \lambda}\frac{d^2A_1}{dt^2}$. In addition, using Eq.~(\ref{ysh-exp}), we find  the volume difference as a function of time, $v_{\text{du}}(t)=2\int_0^1 y_{\text{sh}}(x,t)dx=4A_1(t)/\pi$. Thereafter, we solve Eq.~(\ref{eom-An}) numerically  with $N=2$ and plot the parametric solution ($\bar{p}_{\text{ud}}(t)$,$v_{\text{du}}(t)$) in the range $t\in[0,t_{\text{p}}]$.

\begin{figure}[h!]
\begin{centering}
\includegraphics[height=3.47cm]{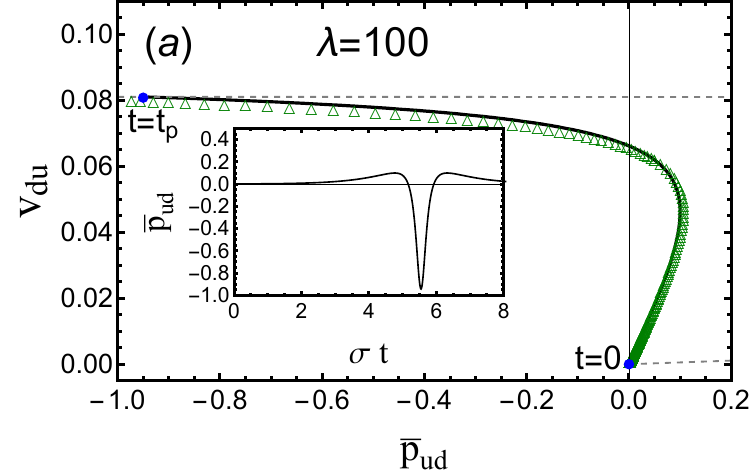} \ \ \  \ \ \ \ \ \includegraphics[height=3.47cm]{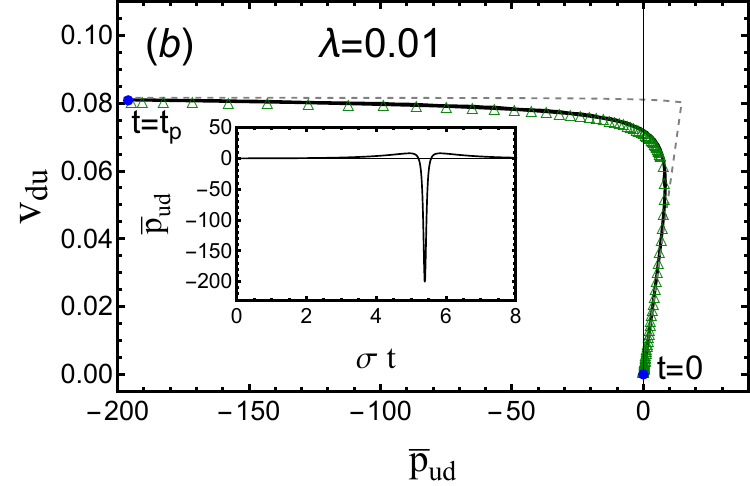} \ \ \  \ \ \ \ \ \includegraphics[height=3.47cm]{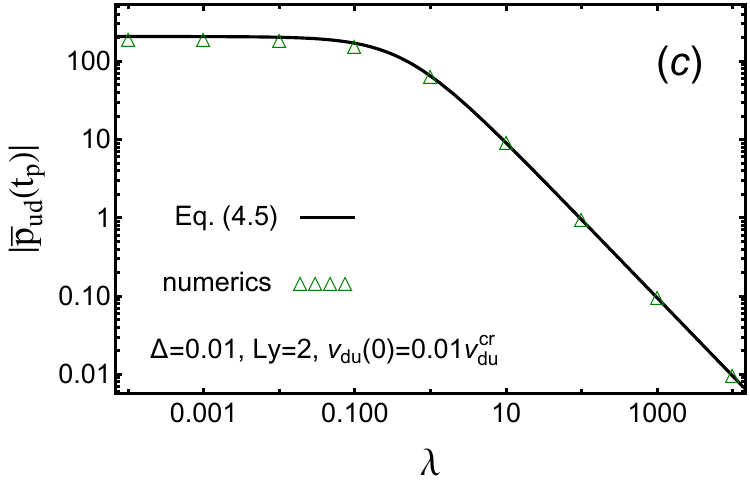} 
%\captionsetup{format=plain,justification=justified}
\caption{The $\bar{p}_{\text{ud}}$-$v_{\text{du}}$ relation and the maximum backward pressure. In all three panels, open green triangles correspond to the numerical solution of Eqs.~(\ref{continuity-benoulli})-(\ref{bc-sheet}), and solid lines correspond to the solution of the two-mode approximation.  The volume difference as a function of the average pressure difference on the sheet is plotted in ({\it a}) the solid-dominated ($\lambda=100$), and ({\it b}) the fluid-dominated ($\lambda=0.01$)  regions. In both panels, the solid black line corresponds to the solution of Eq.~(\ref{eom-An}) with $N=2$, $\Delta=0.01$, $L_y=2$, and $v_{\text{du}}(0)=0.01v_{\text{du}}^{\text{cr}}$. The blue points correspond to times $t=0$ and $t=t_{\text{p}}$. The dashed gray lines correspond to the static solution, Eqs.~(\ref{asym-branch-static-2}) and (\ref{sym-branch-static-2}). The insets show $\bar{p}_{\text{ud}}(t)$ as a function of time, where $\sigma$ is given by Eq.~(\ref{growth-rate-2nd-app}). In the fluid-dominated region, the maximum backward pressure  ($\bar{p}_{\text{ud}}(t_{\text{p}})\simeq -200$) is much larger than that in the solid-dominated region ($\bar{p}_{\text{ud}}(t_{\text{p}})\simeq -1$). In addition, the evolution of the $\bar{p}_{\text{ud}}$-$v_{\text{du}}$ relation in the fluid-dominated region follows the static solution, except for some deviations close to the asymmetric-to-symmetric transition. ({\it c}) The absolute value of the maximum average pressure difference on the sheet.}
\label{pud-vs-vdu-moderate}
\end{centering}
\end{figure}

The results of these solutions are plotted in figures~\ref{pud-vs-vdu-moderate}({\it a}) and \ref{pud-vs-vdu-moderate}({\it b}), for the solid- ($\lambda=100$) and fluid- ($\lambda=0.01$) dominated regions, respectively.  Qualitatively, the two regions exhibit similar behavior. The pressure difference in the chamber increases from almost zero up to a maximum positive value, from which it rapidly decreases and becomes negative. The backward pressure  is maximized at the peak time $t_{\text{p}}$; see the time-dependent behaviour of $\bar{p}_{\text{ud}}(t)$  in the insets of these figures. %The negative pressure difference corresponds to the tendency of the flow to move in the reverse direction, from the down-up to the up-down direction. 
Quantitatively, however, the two profiles are considerably different, because the maximum backward pressure is much larger in the fluid-dominated region (figure~\ref{pud-vs-vdu-moderate}({\it b})), than in the solid-dominated region (figure~\ref{pud-vs-vdu-moderate}({\it a})). 

%The transition from positive to negative pressure differences can be explained as follows: Initially, the system has a negative feedback loop between the sheet and fluid, where the sheet's motion drives the fluid's dynamics, which in turn applies a positive pressure difference and resists the sheet's motion. As the system evolves, the fluid gains kinetic energy and reduces its resistance to the sheet's motion, until the pressure difference reaches zero and the resistance is eliminated. At this point, the pressure difference becomes negative and the system exhibits positive feedback, where the sheet transfers energy to the fluid, which enhances the sheet's motion. This positive feedback continues until the system reaches its maximum volume difference. High energy flows have larger pressure differences than low energy flows.

The transition from positive to negative pressure differences can be explained as follows: Initially, the system exhibits a ``negative feedback'' between the sheet and the fluid, meaning that the sheet's motion drives the fluid's dynamics, which, in turn applies a positive pressure difference and resists the sheet's motion. As the system evolves, the fluid continuously gains kinetic energy and reduces its resistance to the sheet's motion. Then, at some instant, the pressure difference vanishes, $\bar{p}_{\text{ud}}=0$, and the resistance of the fluid is almost eliminated. Beyond this moment, the pressure difference becomes negative, and the system exhibits a ``positive feedback'', meaning that the sheet transfers energy to the fluid, which, in turn enhances the sheet's motion. This positive feedback accelerates the system's dynamics and creates a spike of pressure drop in the chamber. The process terminates when the system meets the constraint on the maximum volume difference.

%As expected in this process, high energy flows ($\lambda\ll 1$) exhibit larger pressure differences than low energy flows ($\lambda\gg 1$).}}
%{\color{blue}{We can explain the transition from positive to negative pressure differences as follows. Initially, the system exhibits a ``negative feedback'' between the sheet and the fluid, i.e., the sheet's motion drives the fluid's dynamics that, in return, applies a positive pressure difference that slows down the sheet. As the system evolves, the fluid continuously gains kinetic energy that reduces its resistance to the sheet's motion until at some instant $\bar{p}_{\text{ud}}=0$, and the resistance approximately vanishes to zero. At this moment, the pressure difference becomes negative, and the system exhibits ``positive feedback'', i.e., the sheet transfers energy to the fluid, that in return, enhances the sheet's motion. This positive feedback terminates when the system meets the constraint on the maximum volume difference. As expected in this process, high energetic flows ($\lambda\ll 1$) exhibit larger pressure differences than low energetic flows ($\lambda\gg 1$). }}

%Indeed, slowing down and reversing the direction of a high energetic flow ($\lambda\ll 1$) requires larger pressure differences than are required for the low energetic flow ($\lambda\gg 1$). 

To estimate the magnitude of the pressure spike, $\bar{p}_{\text{ud}}(t_{\text{p}})$, we use the two-mode approximation.  Recalling that near the peak time when the sheet is close in shape to the first mode of buckling, i.e., $A_1(t_{\text{p}})\simeq 2\Delta^{1/2}/\pi$ and $A_2(t_{\text{p}})\simeq 0$, and that $E\simeq 4\pi^2\Delta$, we find from Eqs.~(\ref{eom-An-2}) and (\ref{two-modes-approx}) an expression for $\frac{d^2A_1}{dt^2}(t_{\text{p}})$ as a function of the system's parameters. This gives the maximum backward pressure,
%In addition, we use Eq.~(\ref{eom-An-2}) and Eq.~(\ref{two-modes-approx}) with $E\simeq 4\pi^2\Delta$, to calculate the second derivative of $A_1(t)$ at $t=t_{\text{p}}$.  Using this result we obtain
\begin{equation}\label{maximum-ave-pud}
\bar{p}_{\text{ud}}(t_{\text{p}})\simeq-\frac{432\pi^5 L_y\Delta^{1/2}}{9\pi^3\lambda+128\tanh(\pi L_y/2)}.
\end{equation}
This analytical approximation compares well with the numerical solution of Eqs.~(\ref{continuity-benoulli})-(\ref{bc-sheet}); see figure~\ref{pud-vs-vdu-moderate}({\it c}). Therefore, in the solid-dominated region, the maximum backward pressure decays to zero as $\bar{p}_{\text{ud}}(t_{\text{p}},\lambda\gg 1)\simeq -48\pi^2 L_y\Delta^{1/2}/\lambda$, whereas  in the fluid-dominated region it is independent of $ \lambda$, i.e., $\bar{p}_{\text{ud}}(t_{\text{p}},\lambda\ll 1)\simeq -(27\pi^5/8)L_y\Delta^{1/2}/\tanh\left(\pi L_y/2\right)$. %{\color{blue}{Why $\bar{p}_{\text{ud}}(t_{\text{p}})$ increases with $L_y$?}} %{\color{blue}{Interestingly, the maximum backward pressure increases with $L_y$ }}

Note also that  the $\bar{p}_{\text{ud}}$-$v_{\text{du}}$ relation approximately follows the static solution, Eqs.~(\ref{asym-branch-static-2}) and (\ref{sym-branch-static-2}), in the fluid-dominated region; see the dashed lines in figure~\ref{pud-vs-vdu-moderate}({\it b}). This is because the dynamics of the fluid in this region is much slower than that in the solid-dominated region. Nonetheless, deviations between the two solutions, static and dynamic, are observed close to the asymmetric-to-symmetric transition. These deviations may be attributed to inertial effects in the dynamics solution.

%\bigskip\bigskip\bigsip

\section{Discussion on experimental consequences}
\label{discussion}

%\subsection{Main results in dimensional form}
%\label{exp-conseq}

To fully define the system, eight physical parameters are needed: Four parameters specify the properties of the sheet $\{ \tilde{L},\tilde{\rho}_{\text{sh}},\tilde{B},\tilde{h} \}$, three parameters define the dimensions of the chamber $\{ \tilde{L}_x,\tilde{L}_y,\tilde{W} \}$, and an additional parameter characterizes the fluid $\tilde{\rho}_{\ell}$. The control parameter is the initial volume difference in the chamber $\tilde{v}_{\text{du}}(0)=\tilde{v}_{\text{d}}(0)-\tilde{v}_{\text{u}}(0)$.

%Eight physical parameters are required to define the system. Four parameters define the sheet $\{\tilde{L},\tilde{\rho}_{\text{sh}},\tilde{B},\tilde{h}\}$, three define the dimensions of the chamber $\{\tilde{L}_x,\tilde{L}_y,\tilde{W}\}$, and another parameter characterizes the fluid $\tilde{\rho}_{\ell}$. The control parameter is the initial volume difference in the chamber $\tilde{v}_{\text{du}}(0)=\tilde{v}_{\text{d}}(0)-\tilde{v}_{\text{u}}(0)$. 

%To facilitate comparisons of the theory with experiments, we repeat two of our central predictions in dimensional form. The first prediction corresponds to the case where the initial configuration of the sheet is closed in shape to the second buckling mode. In this case, the experimentally measurable quantities are the growth rate $\sigma$, which is given in Eq.~(\ref{growth-rate-2nd-app}) and characterizes the early time of the evolution, and the time $t_{\text{p}}$, which is given in Eq.~(\ref{scaling-sigma-tp}) and characterizes the behaviour at moderate times. In dimensional form, these quantities read:
To facilitate comparisons between our theory and experimental observations, we present two of our central predictions in dimensional form. The first prediction relates to the scenario where the initial configuration of the sheet is close in shape to the second buckling mode. In this case, the experimentally measurable quantities are the growth rate $\sigma$, which is defined in Eq.(\ref{growth-rate-2nd-app}) and characterizes the early stage of the evolution, and the time $t_{\text{p}}$, which is given in Eq.(\ref{scaling-sigma-tp}) and characterizes the behaviour at moderate times. In dimensional form, these quantities are given by:
\begin{subequations}\label{exp-sigma-tp}
\begin{eqnarray}
\tilde{\Delta}/\tilde{L}\ll 1\ \ \ &\text{and}& \ \ \  \tilde{v}_{\text{du}}(0)/ \tilde{v}_{\text{du}}^{\text{cr}}\ll 1:  \nonumber \\ 
&&\tilde{\sigma}=\frac{\sqrt{3}\pi^2}{\sqrt{1+\frac{8}{\pi^2}\frac{\tilde{\rho}_{\ell}\tilde{L}_y}{\tilde{\rho}_{\text{sh}} \tilde{h}}}}\left(\frac{\tilde{B}}{\tilde{\rho}_{\text{sh}} \tilde{h} \tilde{L}^4}\right)^{1/2}, \label{exp-sigma-tp-1}\\ 
%&& \tilde{t}_{\text{p}}\simeq \frac{\sqrt{1+\frac{8}{\pi^2}\frac{\tilde{\rho}_{\ell}\tilde{L}_y}{\tilde{\rho}_{\text{sh}} \tilde{h}}}}{\sqrt{3}\pi^2}\ln\left(\frac{\pi\sqrt{3(15+2\pi^2)}}{(3+\pi^2)}\frac{\tilde{v}_{\text{du}}^{\text{cr}}}{\tilde{v}_{\text{du}}(0)}\right)\left(\frac{\tilde{\rho}_{\text{sh}} \tilde{h} \tilde{L}^4}{\tilde{B}}\right)^{1/2}, \\
&& \tilde{\sigma}\tilde{t}_{\text{p}}\simeq \ln\left(\frac{\pi\sqrt{3(15+2\pi^2)}}{(3+\pi^2)}\frac{\tilde{v}_{\text{du}}^{\text{cr}}}{\tilde{v}_{\text{du}}(0)}\right),\label{exp-sigma-tp-2}
\end{eqnarray}
\end{subequations}
where $\tilde{v}_{\text{du}}^{\text{cr}}=\frac{2(3+\pi^2)}{\pi\sqrt{3(15+2\pi^2)}}\tilde{\Delta}^{1/2}\tilde{L}^{3/2}\tilde{W}$ is the volume difference at the asymmetric-to-symmetric transition in the quasi-static solution ($\S$~\ref{recap-static}), and we keep in mind that the normalized excess length is assumed small in our analysis, i.e., $\tilde{\Delta}/\tilde{L}=(\tilde{L}-\tilde{L}_x)/\tilde{L}\ll 1$.

The second prediction of our theory corresponds to the frequency of oscillations $\omega$ around the first buckling mode, Eq.~(\ref{omega-1st-mode-N2}).  In dimensional form the frequency is given by:
\begin{equation}\label{dimensional-omega}
\tilde{\omega}=\frac{2\sqrt{3}\pi^2}{\sqrt{1+\frac{128}{9\pi^3}\tanh\left(\frac{\pi \tilde{L}_y}{2\tilde{L}}\right)\frac{\tilde{\rho}_{\ell}\tilde{L}}{\tilde{\rho}_{\text{sh}} \tilde{h}}}}\left(\frac{\tilde{B}}{\tilde{\rho}_{\text{sh}} \tilde{h} \tilde{L}^4}\right)^{1/2},
\end{equation}
where in the small-amplitude approximation the first buckling mode is obtained when $\tilde{v}_{\text{du}}(0)=\frac{8}{\pi^2}\tilde{\Delta}^{1/2}\tilde{L}^{3/2}\tilde{W}$.  It is important to note that this analytical prediction is valid only for very small excess lengths $\tilde{\Delta}/\tilde{L}\lesssim 0.01$. Larger excess lengths will probably result in significant quantitative deviations from this solution.

% and we keep in mind that a convergence to this analytical prediction is obtained for very small excess lengths $\tilde{\Delta}/\tilde{L}\lesssim 0.01$. %At excess higher lengths we anticipate large quantitative deviations from this solution. 

To obtain a sense of the physical time and pressure scales that can potentially be induced in the system, let us consider a chamber, with dimensions $\tilde{L}_x=\tilde{L}_y=\tilde{W}=5$ mm, that is filled with water ($\tilde{\rho}_{\ell}\simeq 10^3\ \text{kg/m}^3$). In addition, let us assume that the sheet is made of polyethylene terephthalate with Young's modulus $\tilde{E}\simeq 1$ GPa, Poisson's ratio $\nu\simeq 0.3 $, and thickness $\tilde{h}\simeq 0.1$ mm, such that  the bending modulus is $\tilde{B}=\tilde{E}\tilde{h}^3/[12(1-\nu^2)]\simeq 9\times 10^{-5} $J \cite{Gomez2017}. The density of the sheet is approximately $\tilde{\rho}_{\text{sh}}\simeq 1500\ \text{kg/m}^3$, and its total length is $\tilde{L}=5.5$ mm ($\tilde{\Delta}/\tilde{L}=0.09$). Under these conditions, the inertial timescale of the sheet is $\tilde{t}_{\star}=(\tilde{\rho}_{\text{sh}} \tilde{h} \tilde{L}^4/\tilde{B})^{1/2}\simeq 10^{-3}$ s, and the structure-to-fluid mass ratio is given by $\lambda\simeq 0.03$, i.e., the system is in the fluid-dominated region. Since our theory is limited to inviscid fluids, it is reasonable to assume that the theory should agree with the solution of the more general, viscous equations, in the limit of high Reynolds numbers, where energy dissipation is rather small. Estimating the Reynolds number as $\text{Re}\sim\frac{\tilde{\rho}_{\ell}\tilde{\bar{v}}(\tilde{t}_{\text{p}})\tilde{L}}{\tilde{\mu}}$, where $\tilde{\mu}\simeq 10^{-3}\ \text{Pa}\cdot\text{s}$ is the dynamic viscosity, and $\tilde{\bar{v}}(\tilde{t}_{\text{p}})\sim (\lambda \tilde{\Delta}/\tilde{L}_y)^{1/2}(\tilde{L}/\tilde{t}_{\star})$ is our scaling for the fluid's velocity at time $\tilde{t}_{\text{p}}$ in the fluid-dominated region (see $\S$~\ref{elasto-hydrodynamic-energetic-interplay}) we find that $\text{Re}\sim 10^3$, i.e., it is relatively high. Yet, we are aware that higher Reynolds numbers should possibly be considered to reveal a convergence to the inviscid limit.   Using these parameters, we have from Eqs.~(\ref{exp-sigma-tp-1}) and (\ref{dimensional-omega}) that the growth rate  and the frequency of the oscillations are $\tilde{\sigma} \simeq 3.2 /\tilde{t}_{\star} $ and  $\tilde{\omega}  \simeq 8.5 /\tilde{t}_{\star}$, respectively. In addition, given an initial volume difference, the peak time is obtained from Eq.~(\ref{exp-sigma-tp-2}). This gives $\tilde{t}_{\text{p}}\simeq 3.1 \tilde{t}_{\star}$ if $\tilde{v}_{\text{du}}(0)/\tilde{v}_{\text{du}}^{\text{cr}}=10^{-4}$, and $\tilde{t}_{\text{p}}\simeq 1.7 \tilde{t}_{\star}$ if $\tilde{v}_{\text{du}}(0)/\tilde{v}_{\text{du}}^{\text{cr}}=10^{-2}$ . At that moment, the average backward pressure difference on the sheet is given by $\tilde{\bar{p}}_{\text{ud}}(t_{\text{p}})\simeq -160$ KPa;  see Eq.~(\ref{maximum-ave-pud}). %, where we remind the reader that the pressure difference is normalized by $B/L^3$.

However, when attempting to predict the behaviour of an experimental system using our solution, it is important to keep in mind the assumptions made in the formulation. For example, we assumed that the fluid exchange between the two parts of the chamber occurs through the upper and lower walls, $y=\pm L_y/2$, but, in practice, this fluid exchange is likely to occur through a connecting channel, as shown in figure~\ref{schematics}. In this case, the analytical analysis must take into account the geometry of the transition region and its associated pressure drop. For example, we anticipate that a narrow transition channel will slow down the dynamics and decrease the growth rate of the instability. Additionally, we expect that if the pressure drop in the channel is much smaller than $\bar{p}_{\text{ud}}(t_{\text{p}})$, it will have a negligible effect on the dynamics. 
%Yet, when trying to predict the behaviour of an experimental system based on our solution, we must keep in mind the assumptions laid early in the formulation. For example, we assumed that the fluid exchange between the two parts of the chamber occurs through the upper and the lower walls, $y=\pm L_y/2$. However, in practice, this fluid exchange will most probably be realized through a connecting channel, as illustrated in Fig.~\ref{schematics}. In this case, one must account in the analytical analysis for the geometry of the transition region and the pressure drop associated with it. For example, we anticipate that a narrow transition channel will slow down the dynamics and decrease the growth rate of the instability. We also expect that if the pressure drop in the channel is much smaller than $\bar{p}_{\text{ud}}(t_{\text{p}})$, it will have a negligible effect on the dynamics.  }}

\section{Concluding remarks}
\label{conclusions}
We investigated the dynamic interaction between a thin sheet and an inviscid fluid that are confined in a closed rectangular chamber. Our investigation focused on two different regions of the system, the early time evolution, where nonlinear effects are negligible, and the evolution at moderate times, where nonlinearity plays a crucial role in the solution. To analyze the dynamics at $t\ll 1$, we employed a linear stability analysis around the second and first buckling modes. While the second mode is always an unstable state whose highest growth rate is a positive number, the first mode is always stable and yields an imaginary growth rate, i.e., periodic oscillations. In the small-amplitude approximation, $\Delta\ll 1$, we obtained analytical solutions for the highest growth rate and the smallest oscillation frequency, Eqs.~(\ref{growth-rate-2nd-app}) and (\ref{omega-1st-mode-N2}) respectively, which agree well with the numerical solutions. Yet, experimental data is needed to validate these central analytical predictions. To facilitate comparisons with experiments, we repeated these results in dimensional form in \S~\ref{discussion}.

%,
%\begin{subequations}\label{dimensional-sigma-omega}
%\begin{eqnarray}
%\tilde{\sigma}&=&\frac{\sqrt{3}\pi^2}{\sqrt{1+\frac{8}{\pi^2}\frac{\tilde{\rho}_{\ell}\tilde{L}_y}{\tilde{\rho}_{\text{sh}} \tilde{h}}}}\left(\frac{\tilde{B}}{\tilde{\rho}_{\text{sh}} \tilde{h} \tilde{L}^4}\right)^{1/2}, \\
%\tilde{\omega}&=&\frac{2\sqrt{3}\pi^2}{\sqrt{1+\frac{128}{9\pi^3}\tanh\left(\frac{\pi \tilde{L}_y}{2\tilde{L}}\right)\frac{\tilde{\rho}_{\ell}\tilde{L}}{\tilde{\rho}_{\text{sh}} \tilde{h}}}}\left(\frac{\tilde{B}}{\tilde{\rho}_{\text{sh}} \tilde{h} \tilde{L}^4}\right)^{1/2}.
%\end{eqnarray}
%\end{subequations}

Given the chamber’s dimensions, we showed that both $\sigma$ and $\omega$ converge to constants in the solid-dominated region and exhibit  $\lambda^{1/2}$ scaling in the fluid-dominated region. This scaling highlights the effect of the fluid on the sheet's motion. When $\lambda\gg 1$, the elastic forces are primarily balanced by the inertia of the sheet, and the fluid has minimal effect on the dynamics. On the other hand, when $\lambda\ll 1$, the elastic forces are mainly balanced by the hydrodynamic pressure difference, and the sheet's inertia has a limited effect on the dynamics.   
%These scalings reveal the effect of the fluid on the sheet's motion. When $\lambda\gg 1$ the elastic forces are balanced by the inertia of the sheet, and the fluid does not play a role in the dynamics. However, when $\lambda\ll 1$ the elastic forces are balanced by the hydrodynamic pressure difference, and the sheet's inertia does not play a significant role in the dynamics.
%the added mass on the sheet's dynamics. When $\lambda\gg 1$, the added mass diminishes to zero, and the sheet's motion is unaffected by the fluid's pressure difference. However, when $\lambda\ll 1$, the added mass is large, and the fluid's pressure difference controls the dynamics and slows down the sheet's motion. 
The differences between these two regions are further manifested in the eigenfunctions of the linear stability solution. In the solid-dominated region only one mode of the sheet is excited, i.e., the other modes fall to zero as $1/\lambda$, while an infinite number of modes are excited in the fluid-dominated region. While this difference did not affect the system's behaviour at moderate times, since in the leading order the dynamics is governed by the first two modes, we conjecture that it can influence the system's behaviour at $t\gg 1$; namely, beyond moderate times when higher modes have an increasing effect on the dynamics, we would expect the fluid-dominated solution to be less ordered than the solution in the solid-dominated region. 

In addition, we focused on how the sheet escapes from the second buckling mode at moderate times. Key to this analysis is the two-mode approximation that allowed us to analytically analyze the energetic interplay between the sheet and the fluid; see Eq.~(\ref{energies-Ek-Ef}) and figure~\ref{energies}. This energetic interplay is based on the fact that our model incorporates an elastic sheet and an inviscid fluid, which ensures that the system's total energy is always conserved. At each moment in time, the total energy is distributed between the kinetic energy of the sheet, the potential energy of the sheet, and the energy of the fluid, in different proportions. 
%This energetic interplay relies on the fact that our model considers an elastic sheet and an inviscid fluid, so the system's total energy is always conserved. At each moment in time, the total energy is divided differently between the kinetic energy of the sheet, the potential energy of the sheet, and the fluid's energy. }} 
In the solid-dominated region, the sheet's initial potential energy is converted to the sheet's kinetic energy, and only a small fraction, of the order $\lambda^{-1}$, is converted to the energy of the fluid. However, the picture is reversed in the fluid-dominated region, where most energy is used to displace the fluid. 

The time at which the potential energy of the sheet reaches a minimum, Eq.~(\ref{scaling-sigma-tp}), constitutes another central result of our theory, which can be verified experimentally. In particular, we showed that $\sigma t_{\text{p}}$ is almost independent of the parameter $\lambda$, and in the limit $v_{\text{du}}(0)\ll 1$ it diverges logarithmically. % as $\sigma t_{\text{p}}\propto \ln[1/v_{\text{du}}(0)]$. 

During the dynamic evolution at moderate times, we observed that the sheet-fluid interplay experiences a transition from negative to positive feedback. Initially, the fluid resists the sheet's motion, but at later times, the hydrodynamic pressure difference acts in the direction of the sheet's motion and promotes the sheet's dynamics. The positive feedback ends at $t=t_{\text{p}}$, when the volume difference reaches its maximum value, dictated by the inextensibility of the sheet. At that moment, the pressure difference on the sheet, $\bar{p}_{\text{ud}}$, reaches its peak value.

An important extension of the present theory is the inclusion of viscosity in the mathematical formulation of the fluid. This extension will allow us to investigate the behaviour of the sheet at both high and low Reynolds numbers and thus look for the elasto-hydrodynamic instabilities caused by viscous effects. It will also allow us to investigate the formation of boundary layers and examine their impact on the dynamics of the system.  We will pursue this extension in a future study.

\appendix

\section{Derivation of Eq.~(\ref{energy-tot})}
\label{energy-der-app}
Equations~(\ref{continuity-benoulli})-(\ref{bc-sheet}) have a conserved first integral that corresponds to the total energy in the system. To derive this conserved quantity, we multiply Eq.~(\ref{force-balance-1}) by $\partial\theta/\partial t$, and Eq.~(\ref{force-balance-2}) by $\partial {\bf x}_{\text{sh}}/\partial t$, and subtract the second equation from the first. Then, we integrate the resulting equation between $s\in[0,1]$, and use integration by parts and the geometric constraints, Eq.~(\ref{geo-rel}), to simplify the result. This gives,
\begin{equation}\label{energy-der-1}
\frac{d}{dt}\left[E_{\text{sh}}^{\text{k}}(t)+E_{\text{sh}}^{\text{p}}(t)\right]=-\int_0^1\frac{\partial}{\partial s}\left(\frac{\partial {\bf x}_{\text{sh}}}{\partial t}\cdot {\bf F}-\frac{\partial\theta}{\partial t}\frac{\partial\theta}{\partial s}\right)ds-\int_0^1(p_{\text{u}}-p_{\text{d}})\frac{\partial {\bf x}_{\text{sh}}}{\partial t}\cdot {\bf \hat{n}}_\text{d}ds,
\end{equation}
where $E_{\text{sh}}^{\text{k}}(t)=\frac{1}{2}\int_0^1\left|\frac{\partial{\bf x}_{\text{sh}}}{\partial t}\right|^2ds$ and $E_{\text{sh}}^{\text{p}}(t)=\frac{1}{2}\int_0^1\left(\frac{\partial\theta}{\partial s}\right)^2ds$ are readily identified as the kinetic and potential energies of the sheet, respectively. Given the boundary conditions on the sheet's edges, Eq.~(\ref{bc-sheet}), the first term in the right hand-side of Eq.~(\ref{energy-der-1}) vanishes. Therefore, to complete the derivation, it remains to show that the second term in the right-hand side of Eq.~(\ref{energy-der-1}) equals  $-dE_{\text{f}}(t)/dt$. 

Following Ref.~\cite{Lamb},  the kinetic energy of an incompressible fluid is given by, %the fluids in each part of the chamber is given by,
\begin{equation}\label{energy-der-2}
\frac{d}{dt}\left(\frac{1}{2\lambda}\iint_{v_i(t)}\left|\nabla \phi_i\right|^2dxdy\right)=-\oint_{\delta v_i(t)}p_i \nabla\phi_i\cdot {\bf \hat{n}}_id\tilde{s},
\end{equation}
where $\delta v_i(t)$ are the perimeters of the upper or the lower parts of the chamber, $d\tilde{s}$ is an infinitesimal element on $\delta v_i(t)$ (on the sheet $d\tilde{s}=ds$), and ${\bf \hat{n}}_i(\tilde{s},t)$  are the corresponding local unit normal vectors on $\delta v_i(t)$. Since $\left(\nabla\phi_i \cdot {\bf \hat{n}}_i\right)_{x=0,1}=0$ on the sidewalls of the chamber, in accordance with Eq.~(\ref{bc-fluid-chamber-1}), and since we have periodic boundary conditions on the upper and lower walls, %$p_i(x,\pm L_y/2,t)=0$, in accordance with Eq.~(\ref{bc-fluid-chamber-2}),  
the right hand side of Eq.~(\ref{energy-der-2}) reduces to an integral over the configuration of the sheet. When summed over the two parts of the chamber this gives 
\begin{equation}\label{energy-der-3}
\frac{dE_{\text{f}}(t)}{dt}=-\sum_{i=\text{u,d}}\int_{0}^1 p_i \nabla\phi_i\cdot {\bf \hat{n}}_ids=-\sum_{i=\text{u,d}}\int_0^1 p_i \frac{\partial {\bf x}_{\text{sh}}}{\partial t}\cdot {\bf \hat{n}}_ids.
%\frac{d}{dt}\left(\frac{1}{2\lambda}\iint_{v_i(t)}\left|\nabla \phi_i\right|^2dxdy\right)=-\sum_{i=\text{u,d}}\int_{0}^1 p_i \nabla\phi_i\cdot {\bf \hat{n}}_ids=-\int_0^1 p_i \frac{\partial {\bf x}_{\text{sh}}}{\partial t}\cdot {\bf \hat{n}}_ids,
\end{equation}
Here, $E_{\text{f}}(t)=\sum_{i=\text{u,d}}\frac{1}{2\lambda}\iint_{v_i(t)}\left|\nabla\phi_i\right|^2dxdy$ is the energy of the fluid, and in the second equality we used the kinematic boundary condition, Eq.~(\ref{kinematic-bc}), to replace the normal velocity of the fluid with the normal velocity of the sheet. Finally, note that on the sheet the normal vectors are related by, ${\bf \hat{n}}_{\text{u}}(s,t)=-{\bf \hat{n}}_{\text{d}}(s,t)$. Using this relation and Eq.~(\ref{energy-der-3}), we obtain that,
\begin{equation}\label{energy-der-4}
\frac{d E_{\text{f}}(t)}{dt}=\int_0^1(p_{\text{u}}-p_{\text{d}})\frac{\partial {\bf x}_{\text{sh}}}{\partial t}\cdot {\bf \hat{n}}_\text{d}ds.
\end{equation}
%where $E_{\text{f}}=\sum_{i=\text{u,d}}\frac{1}{2\lambda}\iint_{v_i(t)}\left|\nabla\phi_i\right|^2dxdy$ is the energy of the fluids. 
Substituting Eq.~(\ref{energy-der-4}) into Eq.~(\ref{energy-der-1}) and integrating once with respect to time completes the derivation. 

\section{Minimization of the action}
\label{lag-mini}

In this Appendix, we show that the minimization of the action, $\mathcal{S}=\int_0^T\mathcal{L}dt$ where $\mathcal{L}$ is given by Eq.~(\ref{lagrangian}), yields the complete set of equilibrium equations in the small-amplitude approximation, Eqs.~(\ref{constraint-app})-(\ref{benuolli-kin}). To do so, we minimize the action with respect to the elastic fields, $y_{\text{sh}}(x,t)$ and  $F_x(t)$, and the hydrodynamic fields, $\phi_{\text{u}}(x,y,t)$ and $\phi_{\text{d}}(x,y,t)$,  in the standard way. We consider a small perturbation in each of these variables, for example, $y_{\text{sh}}\rightarrow y_{\text{sh}}+\delta y_{\text{sh}}$,  and then expand the action to linear order in the perturbation, $\delta y_{\text{sh}}(x,t)$.  This procedure gives, after integration by parts, the variation,
\begin{eqnarray}\label{var-L}
\delta\mathcal{S}&=&\int_0^1\left[\left(\frac{\partial y_{\text{sh}}}{\partial t}+\frac{\phi_{\text{d}}(x,0,t)-\phi_{\text{u}}(x,0,t)}{\lambda}\right)\delta y_{\text{sh}}\right]_{t=0}^{T}dx\nonumber \\ &+&\int_0^T\left[-\frac{\partial^2 y_{\text{sh}}}{\partial x^2}\frac{\partial\delta y_{\text{sh}}}{\partial x}+\left(\frac{\partial^3 y_{\text{sh}}}{\partial x^3}+F_x \frac{\partial y_{\text{sh}}}{\partial x}\right)\delta y_{\text{sh}}\right]_{x=0}^{x=1}dt \nonumber \\ &-& \int_0^T\int_0^1\left(\frac{\partial^2 y_{\text{sh}}}{\partial t^2}+\frac{\partial^4 y_{\text{sh}}}{\partial x^4}+F_x(t)\frac{\partial^2 y_{\text{sh}}}{\partial x^2}+\left[p_{\text{u}}(x,0,t)-p_{\text{d}}(x,0,t)\right]\right)\delta y_{\text{sh}}dxdt \nonumber \\&+&\int_0^T\int_0^1\left(\frac{1}{2}\left(\frac{\partial y_{\text{sh}}}{\partial x}\right)^2-\Delta\right)\delta F_x dxdt \nonumber \\&+& \frac{1}{\lambda}\int_0^T\int_0^1\left[\delta\phi_{\text{d}}(x,0,t)-\delta\phi_{\text{u}}(x,0,t)\right]\frac{\partial y_{\text{sh}}}{\partial t}dxdt\nonumber \\ &-&\frac{1}{\lambda}\sum_{i=\text{u,d}}\left(\int_0^T \oint_{\delta v_i} \nabla\phi_i\cdot {\bf \hat{n}}_i \delta\phi_i d\tilde{s}dt -\int_0^T\int_{v_i}\nabla^2\phi_i \delta\phi_i dxdydt\right), %\nonumber \\
\end{eqnarray}
where $v_i$ are the volumes of the chamber above and below the sheet in the small-amplitude approximation, $\delta v_i$ are the perimeters of the upper and lower volumes, ${\bf \hat{n}}_i$ are the unit normal vectors on $\delta v_i$, and $d\tilde{s}$ is an infinitesimal line element on $\delta v_i$ (on the sheet $d\tilde{s}=ds$).

The initial conditions of the system and the boundary conditions that we imposed, Eqs.~(\ref{bc-sheet-2}) and (\ref{bc-sheet-3}), imply that the first and second lines in Eq.~(\ref{var-L}) vanish altogether. The third and fourth lines in Eq.~(\ref{var-L}) vanish if the force balance equation, Eqs.~(\ref{force-balance-app}) and (\ref{benuolli-kin-1}), and the geometric constraint, Eq.~(\ref{constraint-app}), are both satisfied. In the last line, the two integrals over the upper and lower volumes of the chamber, $v_i$, vanish if the continuity equations, Eq.~(\ref{continuity-benoulli-1}), are satisfied. Therefore, it remains to show that the fifth line and the penultimate term in the last line  of Eq.~(\ref{var-L}) are equal to zero. To do so, we note that $\left(\nabla\phi_i\cdot {\bf \hat{n}}_i\right)_{x=0,1}=0$, and that we assumed periodic boundary conditions at $y=\pm L_y/2$, Eq.~(\ref{bc-fluid-chamber-2}). % the perturbation on the upper and lower walls of the chamber vanish, i.e., $\delta \phi_i(x,\pm L_y/2,t)=0$, since the values of the potential functions are fixed at these walls, i.e., $\phi_i(x,\pm L_y/2,t)=0$.  
As a result, the integrals over the perimeters $\delta v_i$ reduce to integrals over the sheet-fluid interfaces. In that case, the remaining part of the variation of $\delta \mathcal{S}$ reads:
\begin{eqnarray}\label{var-L1}
\delta\mathcal{S}&=& \frac{1}{\lambda}\int_0^T\int_0^1\left[\delta\phi_{\text{d}}(x,0,t)-\delta\phi_{\text{u}}(x,0,t)\right]\frac{\partial y_{\text{sh}}}{\partial t}dxdt\nonumber \\ &-&\frac{1}{\lambda}\int_0^T\int_0^1\left[\left(\frac{\partial\phi_{\text{d}}}{\partial y}\delta\phi_{\text{d}}\right)_{y=0}-\left(\frac{\partial\phi_{\text{u}}}{\partial y}\delta\phi_{\text{u}}\right)_{y=0}\right]dxdt.
\end{eqnarray}
Collecting the terms that are proportional to $\delta\phi_i(x,0,t)$, we find that the integrands in Eq.~(\ref{var-L1}) vanish when the kinematic boundary conditions, Eq.~(\ref{benuolli-kin-2}), are satisfied. 

This completes the proof that the force balance equations, and their corresponding boundary conditions in the small-amplitude approximation,  both emanate from the minimization of the action.

\section{Derivation of Eqs.~(\ref{lag-app2}) and (\ref{TVC-matrices})}
\label{lag-derivation}

In this Appendix, we derive Eqs.~(\ref{lag-app2}) and (\ref{TVC-matrices}) in the main text. To do so, we first express the Lagrangian, Eq.~(\ref{lagrangian}), in terms of the unknown time-dependent coefficients, $A_n(t)$,  $a_{m}(t)$, and $c_m(t)$. Substituting the normal mode expansion of the sheet's height function, Eq.~(\ref{ysh-exp}), and the potential functions, Eq.~(\ref{phi-exp}), into the Lagrangian, Eq.~(\ref{lagrangian}), and integrating over the spatial coordinates gives,
\begin{eqnarray}\label{lag11}
\mathcal{L}&=&\sum_{n=1}^{N}\frac{1}{4}\left[\left(\frac{dA_n}{dt}\right)^2+\pi^2 n^2\left(F_x(t)-\pi^2n^2\right)A_n^2\right]-F_x(t)\Delta \nonumber \\ &+&\frac{L_y}{\lambda}a_0\sum_{n=1}^{N}W(n,0)\frac{dA_n}{dt}+\sum_{n=1}^{N}\sum_{m=1}^{N-1}\frac{2}{\lambda}W(n,m)\sinh\left(\frac{\pi m L_y}{2}\right)(a_m-c_m)\frac{dA_n}{dt} \nonumber \\
&-&\frac{L_y}{2\lambda}a_0^2-\sum_{m=1}^{N-1}\frac{\pi m}{2\lambda}\sinh(\pi m L_y)\left(a_m^2+c_m^2\right),
\end{eqnarray} 
While the first line in this equation describes the kinetic and the potential energies of the sheet and the geometric constraint,  the second and third lines emanate, respectively, from the mixed term, $\phi_i(x,0,t)\partial y_{\text{sh}}/\partial t$,  and the kinetic energies of the fluid. 

The next step is to express the coefficients of the hydrodynamic potentials, $a_m(t)$ and $c_m(t)$, in terms of the elastic coefficients, $A_n(t)$. Minimizing Eq.~(\ref{lag11}) with respect to  $a_m(t)$ and $c_m(t)$, we obtain,
\begin{subequations}\label{am-sol}
\begin{eqnarray}
a_0(t)&=&\sum_{k=1}^N W(k,0)\frac{dA_k}{dt}, \label{am-sol-1}\\
a_m(t)&=&-c_m(t)=\sum_{k=1}^N\frac{2}{\pi m}\frac{\sinh(\pi m L_y/2)}{\sinh(\pi m L_y)}W(k,m)\frac{dA_k}{dt},  \ \ \ \ \ \ (m=1,2,...,N-1),\nonumber \\ \label{am-sol-2}
\end{eqnarray}
\end{subequations}
where $W(n,m)= \frac{n}{\pi}\frac{1-(-1)^{n+m}}{n^2-m^2}$ for $n\neq m$ and zero otherwise, as is defined immediately following Eq.~(\ref{TVC-matrices}). Finally, we substitute Eq.~(\ref{am-sol}) back into the Lagrangian, Eq.~(\ref{lag11}), and collect together terms that are proportional to  $\frac{dA_n}{dt}\frac{dA_k}{dt}$, the lateral compression $F_x(t)$, and $A_n A_k$. This yields Eqs.~(\ref{lag-app2}) and (\ref{TVC-matrices}) in the main text.

\section{Linear stability analysis at a finite excess length }
\label{LS-finite-Delta}

When the excess length of the sheet compared to the lateral dimension of the chamber is finite, rather than $\Delta\ll 1$, as assumed in $\S$~\ref{small-slope-formulation}, the linear stability analysis is obtained from the linearization of Eqs.~(\ref{continuity-benoulli})-(\ref{bc-sheet}). In this Appendix, we obtain a closed set of equations for the linearization of the system in this, more general, case and explain the direction we take to obtain the numerical solution.

To linearize  Eqs.~(\ref{continuity-benoulli})-(\ref{bc-sheet}), we first expand the elastic and the hydrodynamic fields around their base solutions; for example, $y_{\text{sh}}(s,t)=y_{\text{sh}}(s,0)+\epsilon e^{\sigma t}\hat{y}_{\text{sh}}(s)$, where $\hat{y}_{\text{sh}}(s)$ is a yet-to-be-determined eigenfunction, and $\epsilon$ is an arbitrary small parameter. Similarly, we define the eigenfunctions $\{\hat{x}_{\text{sh}}(s),\hat{\theta}(s),\hat{F}_x(s),\hat{F}_y(s)\}$ for the elastic sheet, and $\{\hat{\phi}_i(x,y),\hat{p}_i(x,y)\}$ for the fluid. We keep in mind that the fluid starts from rest, and therefore the base solutions for the hydrodynamic fields are equal to zero. Thereafter, we substitute these expansions in the continuity and Bernoulli's equations, Eq.~(\ref{continuity-benoulli}), and expand them to a linear order in $\epsilon$. This expansion reads,
\begin{subequations}\label{hydro-lin-finite-Delta}
\begin{eqnarray}
\nabla^2\hat{\phi}_i&=&0, \\
\hat{p}_i(x,y)&=&-\frac{\sigma}{\lambda}\left[\hat{\phi}_i(x,y)-\hat{\phi}_{\text{d}}\left(\frac{1-\Delta}{2},-\frac{L_y}{2}\right)\right],
\end{eqnarray}
\end{subequations}
where in the last equation we determine the constant $c_i(t)$ such that Eq.~(\ref{ref-pressure}) is satisfied. Similarly, an expansion of the geometric constrains, Eq.~(\ref{geo-rel}), and the force balance equations on the sheet, Eq.~(\ref{force-balance}), gives,
\begin{subequations}\label{elastic-lin-finite-Delta}
\begin{eqnarray}
\frac{d \hat{x}_{\text{sh}}}{d s}&=&-\hat{\theta}\sin\theta(s,0), \\
\frac{d \hat{y}_{\text{sh}}}{d s}&=&\hat{\theta}\cos\theta(s,0),\\
\frac{d^2\hat{\theta}}{d s^2}&=&\left[-F_x(s,0)\hat{\theta}+\hat{F}_y\right]\cos\theta(s,0)-\left[\hat{F}_x+F_y(s,0)\hat{\theta}\right]\sin\theta(s,0),\\
\sigma^2\hat{x}_{\text{sh}}&=&-\frac{d\hat{F}_x}{d s}+\left[\hat{p}_u(x_{\text{sh}}(s,0),y_{\text{sh}}(s,0))-\hat{p}_d(x_{\text{sh}}(s,0),y_{\text{sh}}(s,0))\right]\sin\theta(s,0),\nonumber \\  \\
\sigma^2\hat{y}_{\text{sh}}&=&-\frac{d\hat{F}_y}{d s}-\left[\hat{p}_u(x_{\text{sh}}(s,0),y_{\text{sh}}(s,0))-\hat{p}_d(x_{\text{sh}}(s,0),y_{\text{sh}}(s,0))\right]\cos\theta(s,0). \nonumber \\ 
\end{eqnarray}
\end{subequations}
Equations~(\ref{hydro-lin-finite-Delta}) and (\ref{elastic-lin-finite-Delta}) form a closed system of equations once they are supplemented with the linearized form of the boundary conditions, Eqs.~(\ref{bc-fluid-chamber}), (\ref{kinematic-bc}) and (\ref{bc-sheet}). While at the fluid-chamber and the fluid-sheet interfaces we have,
\begin{subequations}\label{bc-fluid-chamber1}
\begin{eqnarray}
\frac{\partial \hat{\phi}_i}{\partial x}(0,y)&=&\frac{\partial \hat{\phi}_i}{\partial x}(1-\Delta,y)=0,  \\
%\hat{p}_{\text{d}}(x,-L_y/2)&=&\hat{p}_{\text{u}}(x,L_y/2)=0,  \\
\hat{\phi}_{\text{d}}(x,-L_y/2)&=&\hat{\phi}_{\text{u}}(x,L_y/2), \\ 
\frac{\partial\hat{\phi}_{\text{d}}}{\partial y}(x,-L_y/2)&=&\frac{\partial\hat{\phi}_{\text{u}}}{\partial y}(x,L_y/2),  \\ 
\sigma \hat{y}_{\text{sh}}+\frac{\partial\hat{\phi}_i}{\partial x}\left[\frac{\partial y_{\text{sh}}}{\partial x}(s,0)\right]&=&\frac{\partial \hat{\phi}_i}{\partial y},
\end{eqnarray}
\end{subequations}
at the edges of the sheet, the boundary conditions are,
\begin{subequations}\label{bc-sheet-lin}
\begin{eqnarray}
&&\hat{x}_{\text{sh}}(0)=0, \ \ \ \ \ \hat{x}_{\text{sh}}(1)= 0,  \label{bc-sheet-1-lin}  \\
&&\hat{y}_{\text{sh}}(0)=0,  \ \ \ \ \ \hat{y}_{\text{sh}}(1)=0, \label{bc-sheet-2-lin}  \\
&&\frac{d \hat{\theta}}{d s}(0)=0,\ \ \ \ \ \frac{d \hat{\theta}}{d s}(1)=0.\label{bc-sheet-3-lin} %y_{\text{sh}}(0,t)&=&0, \ \ \ \ \ y_{\text{sh}}(1,t)=0, \label{bc-sheet-2}\\
%\partial_s\theta(0,t)&=&0, \ \ \ \ \ \partial_s \theta(1,t)=0, \label{bc-sheet-3}
\end{eqnarray}
\end{subequations}
This completes the linearization of Eqs.~(\ref{continuity-benoulli})-(\ref{bc-sheet}). The linearized equations, Eqs.~(\ref{hydro-lin-finite-Delta})-(\ref{bc-sheet-lin}), always admit the trivial solution, where the eigenfunctions vanish altogether, unless their determinant is equal to zero. 

To solve this set of equations for given $L_y$, $\Delta$, and $\lambda$, we first obtain  numerically the base solution for the position of the sheet, i.e., ${\bf x}_{\text{sh}}(s,0)$ and $\theta(s,0)$. Then, we substitute this solution into the linearized equations and discretize them. The discrete equations are solved using a finite-difference scheme for the elastic sheet and a finite-element scheme for the solution of Eq.~(\ref{hydro-lin-finite-Delta}) in the bulk of the fluid.

%\backsection[Acknowledgements]{We thank Oriel Shoshani and Yoav Green for helpful discussions.}
%
%
%
%\backsection[Funding]{This research was partially supported by the Israel Science Foundation (grant No. 950/22). }

%\bibliographystyle{jfm}
\bibliography{bibitem}{}

\end{document}